\documentclass[aps,twocolumn,pre,showpacs,longbibliography]{revtex4-1}
\usepackage{graphicx}
\usepackage{amsmath}
\usepackage{amsfonts}
\usepackage[table]{xcolor}
\usepackage[caption=false]{subfig}
\usepackage{array}
\usepackage{floatrow}
\usepackage{pbox}
\usepackage{physics}
\usepackage[colorlinks=true,linkcolor=blue,citecolor=blue,filecolor=cyan,urlcolor=blue,breaklinks=true]{hyperref}
\usepackage{mhchem}
\usepackage{braket}

\begin{document}
	\title{Coherence of oscillations in the weak-noise limit}
	
	\author{Benedikt Remlein, Volker Weissmann, and Udo Seifert}
	\address{ II. Institut f\"ur Theoretische Physik, Universit\"at Stuttgart,
		70550 Stuttgart, Germany}
	\date{\today}
	
	\parskip 1mm

	\begin{abstract}
	In a noisy environment, oscillations loose their coherence which can be
	characterized by a quality factor. We determine this quality factor for
	oscillations arising from a driven Fokker-Planck dynamics along a periodic
	one-dimensional potential analytically in the weak noise limit. With this
	expression, we can prove for this continuum model the analog of an upper bound
	that has been conjectured for the coherence of oscillations in
	discrete Markov network models. We show that our approach can also be adapted
	to motion along a noisy two-dimensional limit cycle. Specifically, we
	apply our scheme to the noisy Stuart-Landau oscillator and the thermodynamically
	consistent Brusselator as a simple model for a chemical clock. Our 
	approach thus complements the fairly sophisticated extant general framework
	based on techniques from Hamilton-Jacobi theory with which we compare our
	results numerically.
	\end{abstract}
	
	
	\maketitle

\section{Introduction}

Oscillations are ubiquitous in living systems. The cell cycle \cite{ferr11}, circadian rhythms \cite{naka05,dong08}, glycolysis \cite{gold97}, and biochemical oscillations in general \cite{nova08} represent a few examples for such periodic behavior. These models describe generic processes that are able to keep track of time, hence, to function as clocks. Typically, such biochemical clocks work in surroundings with large fluctuations. Thus, the question arises how these oscillations emerge and how they maintain their coherence. 

On the macroscopic level, bifurcation theory is used to address these questions, i.e., to rationalize the dynamical behavior of deterministic chemical rate equations \cite{nico78, stro00}. A prominent example is Selkov's model for the self-sustained oscillations in glycolysis \cite{selk68}. For microscopic systems, chemical master equations and Markov networks \cite{mcqu67,vankampen,webe17} are established tools to examine these oscillations \cite{gasp02,gasp02a,gonz02, qian06, more07, deys13, bara17,nguy18}. The latter lead to the notion of "Brownian clocks" \cite{bara16,potv16} and to central results like the thermodynamic uncertainty relation in stochastic thermodynamics \cite{bara15,ging16,horo20}, which links the precision of such clocks with the ultimate cost, i.e., the entropy production, see also \cite{piet21}. The first correction of the macroscopic picture towards the microscopic scale can be determined through the weak noise limit \cite{gasp02,gasp02a}. On this mescoscopic scale, the master equation is approximated through a Kramers-Moyal expansion \cite{vankampen}. Thus, the behavior of the oscillations can be studied in the limit where the volume of the system becomes infinite. A recent approach uses large deviation theory to obtain the probability densities beyond the Gaussian approximation \cite{frei21}. An alternative route to examines these oscillations is provided by the phase-reduction method \cite{yosh08,junn09,deys13}.

The macroscopic theory is deterministic, hence, oscillations stay synchronized for all times. For phase diffusion and thus, loss of coherence, randomness in the systems is needed. Cao and co-workers demonstrated for several stochastic models that the number of coherent oscillations increases with increasing entropy production rate \cite{cao15}. This quantity is closely related to the quality factor which we will discuss in detail below. The latter has recently also been used as order parameter to study non-equilibrium phase transitions \cite{nguy18}.

While the mesoscopic theory \cite{gasp02,gasp02a} is rather involved in its general form, we simplify it by performing an explicit calculation for a one-dimensional Fokker-Planck dynamics. Using a spectral decomposition, we derive an original, simple, analytical expression for the quality factor for the motion along this periodic ring.

Such a one-dimensional system can be understood as the continuum limit of a unicyclic Markov network. The quality factor for a Markov network is bounded through its topology and the thermodynamic force driving the system out of equilibrium \cite{bara17}. We will use the relation between a Fokker-Planck dynamic and master dynamics to derive and prove a continuous version of this bound. Exploiting the effective description of a one-dimensional motion in the plane, we show that the number of coherent oscillations of a noisy limit cycle in two dimensions is also given by that of a one-dimensional Fokker-Planck system under the assumption that tangent and normal motion decouple in the weak noise limit. We illustrate our results through three numerical case studies examining a one-dimensional system subject to a periodic force, the noisy Stuart Landau oscillator also known as cubic normal form of a Hopf bifurcation, and the noisy rate equations of the Brusselator in two dimensions. The coherence resulting from the numerics agrees well with our analytical results both for the one-dimensional case and in two-dimensional systems with weak coupling between tangent and normal motion.

This paper is organized as follows. In Sec. \ref{sec:robust}, we discuss the quality factor as a quantitative measure for the coherence of noisy oscillations. Our central result, an explicit expression for the quality factor in one-dimensional Fokker-Planck systems is derived in Sec. \ref{sec:1D}. Using this form, we generalize the microscopic bound \cite{bara17} to dynamics on a one-dimensional ring in Sec. \ref{sec:bound} and illustrate these results numerically in Sec. \ref{sec:examples1}. In Sec. \ref{sec:2D}, we show that oscillations arising from a two-dimensional limit cycle can also be characterized by the one-dimensional expression for the quality factor in the limit of weak noise. We examine the noisy Stuart-Landau oscillator in Sec. \ref{sec:examples2} and discuss in detail the validity of the one-dimensional approximation in Sec. \ref{App:Stuart}. As a generic example for a chemical reaction network, we numerically treat the Brusselator in Sec. \ref{sec:Brusselator} and conclude in Sec. \ref{sec:conclusion}.

\section{Coherence of oscillations}
\label{sec:robust}

A standard measure to describe the coherence of stochastic oscillations is the quality factor $\mathcal R$ \cite{gasp02,gasp02a,bagh14,bara17,nguy18}. This quantity is related to the correlation function of an observable $O(t)$,
\begin{equation}
C_O(t) \equiv \braket{O(t)O(0)}~,
\end{equation}
which displays damped oscillations if $O(t)$ performs a noisy periodic motion. The quality factor,
\begin{equation}
	\mathcal R \equiv \bigg |\frac{\mathrm{Im}\lambda}{\mathrm{Re}\lambda}\bigg |	 = \omega \tau~,
	\label{eq:R}
\end{equation} 
is determined by the dominant eigenvalue 
\begin{equation}
	\lambda = - 1/\tau +i \omega
\end{equation}
of the corresponding dynamical equation. 
The larger $\mathcal R$ is, the longer keeps the oscillator the coherence.

For unicyclic Markov networks, it has recently been conjectured that the quality factor is bounded by
\begin{equation}
\begin{split}
\mathcal R \leq f(\mathcal A,N) \equiv\cot(\pi/N)\tanh(\mathcal A/2N)~,
\end{split}
\label{eq:BS}
\end{equation}
where $N$ is the number of states. The affinity
\begin{equation}
	\mathcal A \equiv \sum_{i=1}^N \ln \frac{k_i^+}{k_i^-}
	\label{eq:Affinity}
\end{equation}
is a function of the rates $k_i^+$ and $k_i^-$ at which a transition from state $i$ to $i \pm 1$ occurs and a measure for the non-equilibrium driving in the network. Eq. (\ref{eq:BS}) is based on strong numerical evidence and covers in its most general form also multicyclic networks \cite{bara17}.

The quality factor for systems following a Fokker-Planck dynamics is obtained by a spectral decomposition of the Fokker-Planck operator \cite{gasp02,gasp02a}. Thus, the number of coherent oscillations is also given by the dominant eigenvalue, Eq. (\ref{eq:R}). This expansion in the eigenbasis leads to a Hamilton-Jacobi equation with the Freidlin-Wentzell Hamiltonian respectively Onsagar-Machlup Lagrangian for the leading order term of the probability density in the weak noise limit \cite{gasp02,gasp02a,bagh14}. The quality factor is then obtained using techniques from Hamilton-Jacobi theory.

We next perform this eigenfunction expansion to find an analytical expression for the quality factor of one-dimensional systems in the weak noise limit without explicitly using Hamilton-Jacobi theory.

\section{One-dimensional Fokker-Planck systems in the weak noise limit}

\subsection{Quality factor}
\label{sec:1D}

Consider a single continuous degree of freedom $x(t)$, e.g., a particle or the concentration of a species on a one-dimensional ring of length $L>0$. It is driven by a spatially periodic force $F(x) = F(x+L)$ and subject to a space-dependent diffusion $Q(x) = Q(x+L)$. We implement the weak noise explicitly through an external parameter $\Omega >0$, which we assume to be large compared to all other system scales, i.e., $\Omega \gg 1$. The dynamics of such a system is then described by the following Fokker-Planck equation \cite{gasp02}
\begin{equation}
	\begin{split}
	\partial_t p(x,t) &= - \{\partial_x[F(x) p(x,t)] - \frac{1}{\Omega}\partial^2_x[Q(x)p(x,t)]\}~\\
	& \equiv \mathcal L_xp(x,t)	~,
\end{split}
	\label{eq:FP1}
\end{equation}
where we assume that $Q(x)>0$ for all $x$. $\mathcal L_x$ denotes the Fokker-Planck operator. Additionally, we require that the force has a unique sign. The latter assumptions ensures that the dynamics has no rest position, which would destroy oscillations in the weak noise limit. Without loss of generality, we assume $F(x)$ to be positive.

Following \cite{gasp02}, we expand Eq. (\ref{eq:FP1}) to obtain the eigenvalues and thus, the quality factor. With the splitting ansatz,  $p(x,t) = f(t)h(x)$, Eq. (\ref{eq:FP1}) becomes
\begin{equation}
	\frac{df(t)/dt}{f(t)} = \frac{\mathcal L_x h(x)}{h(x)} \equiv \lambda~.
	\label{eq:EV}
\end{equation}
The time dependence of the probability density is thus given by an exponential $f(t) \propto \exp(\lambda t)$. 

In order to solve for the eigenvalues $\lambda$, we employ the ansatz $h(x) = \exp[-\Omega \phi(x)]$ \cite{gasp02}. Plugging this form into Eq. (\ref{eq:EV}), we obtain an equation for $g(x) \equiv \phi^\prime(x)$,
\begin{equation}
\begin{split}
	F(x)g(x) + Q(x)g(x)^2 + \frac{1}{\Omega^2} Q^{\prime\prime}(x)\\
	- \frac{1}{\Omega}[\lambda + F^\prime(x) + 2 Q^\prime(x)g(x) + Q(x)g^\prime(x)] = 0.
\end{split}
\end{equation}
We solve this relation by expanding $g(x)$ in inverse powers of $\Omega$, i.e.,
\begin{equation}
	g(x) = g_0(x) + \frac{1}{\Omega} g_1(x) + \frac{1}{\Omega^2} g_2(x) + \mathcal O\left(\frac{1}{\Omega^{3}}\right), ~~\Omega \to \infty~.
	\label{eq:pert}
\end{equation}
Comparing the coefficients, we find the following conditions. In $\mathcal O(1)$,
\begin{equation}
F(x)g_0(x) + Q(x)g_0(x)^2 = 0~;
\label{eq:gSystemO0}
\end{equation}
in $\mathcal O(1/\Omega)$,
\begin{equation}
	\begin{split}
-\lambda F(x)g_1(x) + 2Q(x)g_0(x)g_1(x) - F^\prime(x)\\ - Q(x)g_0^\prime(x) - 2g_0(x)Q^\prime(x)= 0~;
\end{split}
	\label{eq:gSystemO1}
\end{equation}
and in $\mathcal O(1/\Omega^2)$,
\begin{equation}
	\begin{split}
F(x)g_2(x) + Q(x)[g_1(x)^2 + 2g_0(x)g_1(x)]\\- Q(x)g_1^\prime(x) - 2g_1(x)Q^\prime(x) + Q^{\prime\prime}(x) = 0~.
	\end{split}
	\label{eq:gSystemO2}
\end{equation}

There are two solutions for the $\mathcal O(1)$ constraint, i.e. $g_0(x) = 0$ and $\tilde g_0(x) = -F(x)/Q(x)$. We neglect the latter since this solution leads to a vanishing quality factor in the weak noise limit, see Appendix \ref{App:Neglect}.

Solving the system of equations above, Eq. (\ref{eq:gSystemO0}-\ref{eq:gSystemO2}), we obtain the leading order terms for $g(x)$,
\begin{equation}
		g_0(x)= 0 ~,
\end{equation}
\begin{equation}
		g_1(x) = \frac{\lambda + F^\prime(x)}{F(x)}~,
\end{equation}
and
\begin{equation}
	\begin{split}
		g_2(x) &= \frac{1}{F(x)^3}[ - \lambda^2 Q(x) - 3 \lambda Q(x) F^\prime(x) - 2 Q(x)F^\prime(x)^2 \\ & \quad + 2 \lambda F(x)Q^\prime(x) + 2 F(x)F^\prime(x)Q^\prime(x)\\ & \quad + F(x) Q(x)F^{\prime\prime}(x) - F(x)^2Q^{\prime\prime}(x)]~.
	\end{split}
\end{equation}
The periodicity, $p(x,t) = p(x+L,t)$, implies that $\phi(x) = \int^x g(u)du$ must satisfy
\begin{equation}
	i 2 \pi k = \Omega\int_0^L dx g(x) = \Omega[\phi(L)-\phi(0)]
	\label{eq:Periodicity}
\end{equation}
for an arbitrary $k\in \mathbb Z$. Plugging the perturbative solution $g(x)$, Eq. (\ref{eq:pert}), into this relation, we can solve the quadratic equation for the eigenvalues as
\begin{equation}
\begin{split}
	\lambda_{\pm}^{(k)} &= \frac{\Omega}{2\int_0^LQ(x)/F(x)^3dx}~\bigg\{ \int_0^L1/F(x)dx \\ &\quad  - \frac 1\Omega \int_0^L Q(x)F^\prime(x)/F(x)^3dx \\ &\quad \pm \bigg[\bigg(\int_0^L1/F(x)dx - \frac 1\Omega \int_0^L Q(x)F^\prime(x)/F(x)^3dx\bigg)^2 \\ &\quad  - i8\pi k\int_0^LQ(x)/F(x)^3dx/\Omega \bigg]^{\frac 12}  \bigg\}~.
\end{split}
\end{equation}

We find the well-behaved solution by a Taylor expansion for $\Omega \gg 1$ as
\begin{equation}
\begin{split}
		\lambda^{(k)} \equiv \lambda_+^{(k)}&= - (2 \pi k)^2 \frac{1}{\Omega} \frac{\int_0^L Q(x)/F(x)^3 dx}{[\int_0^L 1/F(x) dx]^3} + \mathcal O(1)\\
		&\quad  +i~2\pi k \frac{1}{\int_0^L 1/F(x) dx} + i \mathcal O(\Omega)~.
\end{split}
\label{eq:lambda}
\end{equation}
The corresponding eigenfunctions are  given by
\begin{equation}
	\begin{split}
		p^{(k)}(x,t) &= \mathcal N^{(k)} \exp[\lambda^{(k)} t - \Omega \int^x g^{(k)}(u)du]\\
		& \equiv \exp[\lambda^{(k)} t] h^{(k)}(x)
	\end{split}
\end{equation}
where $g^{(k)}(x)$ denotes the solution of Eq. (\ref{eq:pert}) for the eigenvalue $\lambda^{(k)}$ and $h^{(k)}(x)$ the corresponding eigenfunction of the Fokker-Planck operator, Eq. (\ref{eq:EV}).

The second solution for the eigenvalues is 
\begin{equation}
\begin{split}
	\tilde \lambda^{(k)} \equiv \lambda_-^{(k)} &\approx \Omega [\int_0^L 1/F(x) dx]/[\int_0^L Q(x)/F(x)^3dx] \\ &\quad\quad- i 2\pi k /\int_0^L 1/F(x) dx
\end{split}
\end{equation} 
which we can neglect since the $\tilde \lambda^{(k)}$ would lead to a vanishing number of oscillations in the weak noise limit. 

Thus, for a single continuous degree of freedom following a Fokker-Planck dynamics, Eq. (\ref{eq:FP1}), the quality factor is given by
\begin{equation}
\mathcal R_{\mathrm{cont}} \equiv \bigg |\frac{\Im \lambda^{(1)}}{\Re \lambda^{(1)}}\bigg|=\frac{\Omega}{2\pi}\frac{\left( \int_{0}^L \frac{1}{F(x)}dx\right)^2}{\int_0^L \frac{Q(x)}{F(x)^3}dx}~.
\label{eq:R1D}
\end{equation}
This transparent expression is the first main result of this paper. In the next section, we discuss implications of this result for the microscopic bound, Eq. (\ref{eq:BS}).

\subsection{Bound on the quality factor}
\label{sec:bound}

We first calculate a continuum limit  of the bound on the quality factor, Eq. (\ref{eq:BS}), and then use our main result, Eq. (\ref{eq:R1D}), to derive and prove the continuous version of this bound in the form 
\begin{equation}
	\mathcal R_{\mathrm{cont}} \leq f_{\mathrm{cont}}\equiv \frac{\Omega}{2\pi}\int_0^L\frac{F(x)}{Q(x)}dx~.
	\label{eq:BScont}
\end{equation}  
We start by discretizing the Fokker-Planck equation (\ref{eq:FP1}) in space to obtain a master equation for a unicyclic network,
\begin{equation}
	\partial_t p_i(t) = -(k_i^+ + k_i^-)p_i(t) + k_{i+1}^-p_{i+1} + k_{i-1}^+p_{i-1}(t)~,
	\label{eq:master}
\end{equation}
where $p_i(t)$ denotes the probability to be in state $i$ after time $t$ and $k_i^\pm$ are the associated transition rates. To this end, we use central differences to approximate the derivatives, i.e., 
\begin{equation}
\begin{split}
	f'(x) &\approx [f(x+\Delta x) - f(x- \Delta x)]/(2\Delta x)~~~\text{and}\\f''(x) &\approx [f(x+\Delta x) - 2 f(x) + f(x-\Delta x)]/\Delta x^2~. 	
\end{split}\end{equation}
Comparing the result with Eq. (\ref{eq:master}), we get the rates 
\begin{equation}
	\begin{split}
		k_{x-\Delta x}^+ & = \frac{F_{x-\Delta x}}{2 \Delta x} + \frac{Q_{x-\Delta x}}{\Omega\Delta x^2}~,\\
		k_{x+\Delta x}^- & = -\frac{F_{x+\Delta x}}{2 \Delta x} + \frac{Q_{x+\Delta x}}{\Omega\Delta x^2}~.
	\end{split}
\end{equation}
with $F_x \equiv F(x)$,  $\Delta x \equiv L/N$, and relabeling of the states $i \to x = i \Delta x$.

The cycle affinity $\mathcal A$, Eq. (\ref{eq:Affinity}), becomes by a Taylor expansion
\begin{equation}
	\mathcal A = \sum_{x=1}^N \ln\frac{k_x^+}{k_x^-} = \Omega \sum_{x=1}^N \frac{F_x}{Q_x}\Delta x + \mathcal O (\Delta x^2)~.
\end{equation}
 Thus, we find for the bound on the quality factor, Eq. (\ref{eq:BS}), in the continuum limit, i.e., $N \gg 1$ respectively $\Delta x \ll 1$, 
\begin{equation}	
\begin{split}
	f  &= \cot\frac{\pi}{N}\tanh\frac{\mathcal A}{2N} \approx \frac{\Omega}{2\pi} \sum_{x=1}^N\frac{F_x}{Q_x}\Delta x\\ &\approx\frac{\Omega}{2\pi}\int_0^L \frac{F(x)}{Q(x)}dx \equiv f_{\mathrm{cont}}~.
\end{split}
\label{eq:fmac}
\end{equation}

The quality factor $\mathcal R_{\mathrm{cont}}$ from Eq. (\ref{eq:R1D}) and $f_{\mathrm{cont}}$ can be related through a Cauchy Schwartz inequality, 
\begin{equation}
\begin{split}
			\left( \int_0^L \frac{1}{F(x)} dx\right)^2 & = \left( \int_0^L \sqrt{\frac{F(x)}{Q(x)}}\sqrt{\frac{Q(x)}{F(x)^3}} dx\right)^2\\ &\leq \left( \int_0^L \frac{F(x)}{Q(x)} dx\right) \left( \int_0^L \frac{Q(x)}{F(x)^3} dx\right)~.
\end{split}
\end{equation}
Dividing both sides by $\int_0^L Q(x)/F(x)^3dx$ leads to Eq. (\ref{eq:BScont}). Thus, we have proven the continuum version of the bound, Eq. (\ref{eq:BS}), which has been conjectured in \cite{bara17} for discrete unicyclic Markov networks. This is our second main result.

\subsection{A specific example}
\label{sec:examples1}
For an illustration, we consider a particle on a one-dimensional ring of length $L = 2 \pi$. The particle experiences a periodic force $F_\varepsilon(x) \equiv 1 + \varepsilon \sin x$ with $-1 <\varepsilon< 1$. We keep the diffusivity constant, i.e., $Q(x) \equiv 1$. The dynamics is governed by the Langevin equation
\begin{equation}
	\dot x(t) = F_\varepsilon(x(t)) + \sqrt{\frac{2}{\Omega}}\xi(t)
	\label{eq:ex1}
\end{equation}
with $\braket{\xi(t)} = 0$ and $\braket{\xi(t)\xi(t^\prime)} = \delta(t-t^\prime)$.  This equation is equivalent to a Fokker-Planck equation of the form of Eq. (\ref{eq:FP1}). Thus, we can apply the theory developed in the previous sections.

We keep the initial value constant, $x_0 = -\pi$, thus, the correlation function is proportional to the mean value of $x(t)$,
\begin{equation}
	C_x(t) = \braket{x(t)|x(0)=x_0}x_0~. 
\end{equation}
In the weak-noise limit, the deterministic solution for the mean value is given by
\begin{equation}
	\begin{split}
		x_{\infty}(t)  & = -2 \arctan \bigg\{ \varepsilon - \sqrt{1 - \varepsilon^2}  \\ &\times \tan \bigg[ \frac 12 t \sqrt{1 - \varepsilon^2}  + 	\arctan \frac{\varepsilon +\tan(x_0/2)}{\sqrt{1 - \varepsilon^2}}  \bigg] \bigg\}~.
	\end{split}
	\label{eq:1DCor}
\end{equation}
The quality factor for this system can be calculated analytically with Eq. (\ref{eq:R1D}) as
\begin{equation}
	\mathcal R_{\mathrm{cont}} = \Omega \frac{2 (1-\varepsilon^2)^{3/2}}{2 + \varepsilon^2}~.
	\label{eq:R1Dex}
\end{equation}
The bound, Eq. (\ref{eq:fmac}), is simply $f_{\mathrm{cont}} = \Omega$. 

\begin{figure}
	\subfloat[\label{sfig:1a}]{%
		\includegraphics[width=.85\textwidth]{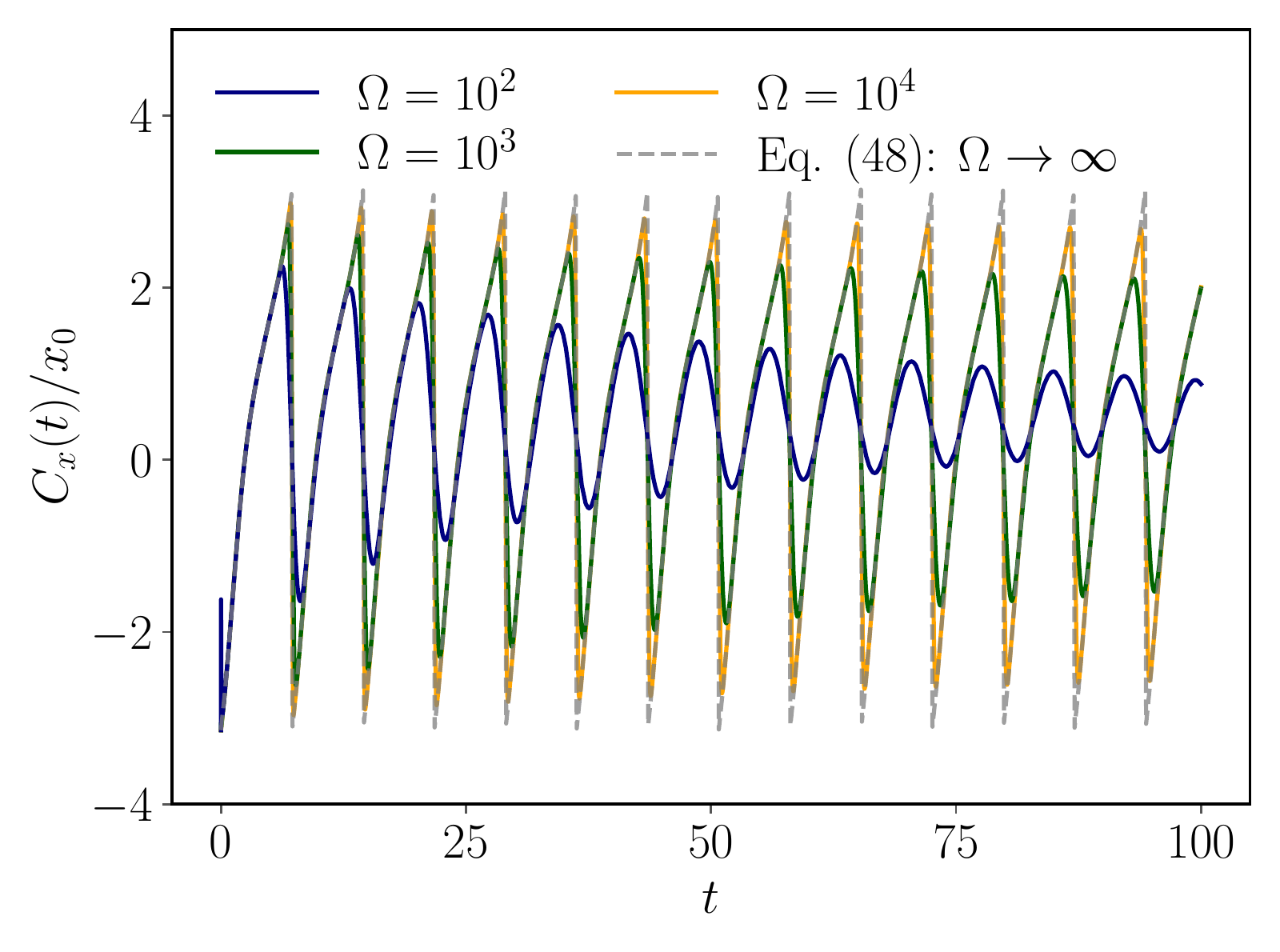}
	}

	\subfloat[\label{sfig:1b}]{%
		\includegraphics[width=.85\textwidth]{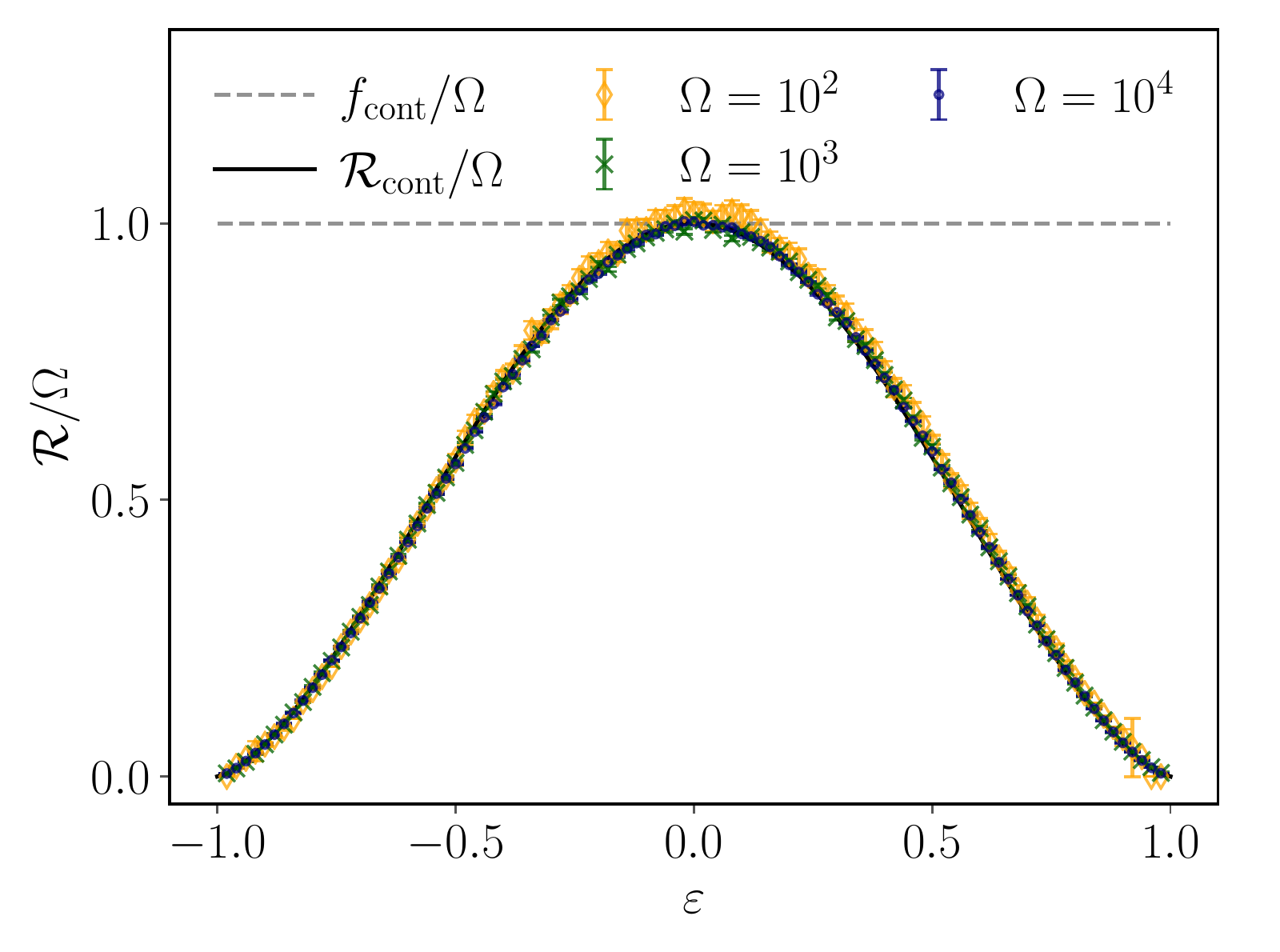}}
	\caption{(a) Correlation function $C_x(t)/x_0$  for various $\Omega$ and $\varepsilon = - 1/2$, $x_0 = -\pi$ for a particle on a ring. (b) Ratio $\mathcal R/\Omega$ as a function of $\varepsilon$. The continuous bound is the gray dashed line. The solid line represents the theoretical prediction for the quality factor, Eq. (\ref{eq:R1Dex}). We include error bars taking into account the fitting error.}
	\label{fig:1}
\end{figure}

As shown in  Fig. \ref{sfig:1a}, we observe a numerical convergence towards the deterministic solution, Eq. (\ref{eq:1DCor}) for increasing $\Omega$, while the phase diffusion smears out the sharp edges the stronger the noise becomes. In this regime, we observe exponentially damped oscillations. From the numerically calculated correlation function, we  obtain the frequency, the decay rate, and the quality factor $\mathcal R$  as given by Eq. (\ref{eq:R}). The ratio $\mathcal R/\Omega$ is shown in Fig. \ref{sfig:1b} for various $\Omega$ as a function of $\varepsilon$. The numerical result for this one-dimensional system agrees very well with our analytical prediction. For vanishing $\varepsilon$, the bound, Eq. (\ref{eq:fmac}), is saturated as predicted by Eq. (\ref{eq:R1Dex}), while the quality factor tends to zero for $\varepsilon \to \pm 1$. This is due to the fact, that the force field, $F_\varepsilon(x) = 1 + \varepsilon \sin x$, establishes a root for $\varepsilon \approx \pm 1$, for which oscillations vanish. 

In what follows, we turn to two-dimensional systems. We demonstrate analytically and numerically that the one-dimensional quality factor, Eq. (\ref{eq:R1D}), is also applicable to planar noisy oscillations for which a limit cycle emerging through a Hopf bifurcation is a prime example.

\section{Planar oscillations in the weak noise limit}
\label{sec:2D}
Consider two continuous degrees of freedom $x(t)$ and $y(t)$. Their coupled dynamics is governed by an autonomous Langevin equation,

\begin{equation}
	\frac{d}{dt} \begin{pmatrix}x(t)\\y(t)\end{pmatrix} = \begin{pmatrix}F_x(x(t),y(t))\\F_y(x(t),y(t))\end{pmatrix} + \frac{1}{\sqrt{\Omega}} 
C(x(t),y(t))\begin{pmatrix}\xi_1(t)\\\xi_2(t)\end{pmatrix}~
	\label{eq:2DLangevin}
\end{equation}
where $\braket{\xi_i(t)}=0$ and $\braket{\xi_i(t)\xi_j(t^\prime)} = \delta_{ij}\delta(t-t^\prime)$. The weak noise limit is again given explicitly through the external parameter $\Omega \gg 1$. We denote the entries of the matrix characterizing the multiplicative noise as 
\begin{equation}
	C(x(t),y(t)) = \begin{pmatrix}c_{x,1}(x(t),y(t)) & c_{x,2}(x(t),y(t))\\c_{y,1}(x(t),y(t))& c_{y,2}(x(t),y(t))\end{pmatrix}~.
	\label{eq:corM}
\end{equation}

We assume that the deterministic dynamics has a stable limit cycle, i.e., a closed curve in phase space which we denote as $\{\bar x, \bar y\}$. Along this curve, the vector-field $F(x,y) \equiv (F_x(x,y),F_y(x,y))$ has no root, thus, a non-vanishing norm
\begin{equation}
\norm{F(\bar x,\bar y)} > 0	~.
\label{eq:norm}
\end{equation}
The norm $\norm{v}$ of a vector $v = (v_x,v_y)$ is induced by the standard scalar-product in $\mathbb R^2$,
\begin{equation}
\norm{v}\equiv \sqrt{v\cdot v} \equiv \sqrt{v_x^2+ v_y^2}~.	
\end{equation}
Continuity of the vector field assures that Eq. (\ref{eq:norm}) holds in a neighborhood of the limit cycle. Thus, we define the vector parallel to the force as
\begin{equation}
	\hat f_{\parallel}(x,y) \equiv  \begin{pmatrix}F_x(x,y)\\F_y(x,y)\end{pmatrix}\bigg /\norm{\begin{pmatrix}F_x(x,y)\\F_y(x,y)\end{pmatrix}}
\end{equation}
and the perpendicular one as
\begin{equation}
	\hat f_{\perp}(x,y) \equiv  \begin{pmatrix}F_y(x,y)\\-F_x(x,y)\end{pmatrix}\bigg /\norm{\begin{pmatrix}F_x(x,y)\\F_y(x,y)\end{pmatrix}}~.
\end{equation}
We denote the projections along these directions as
\begin{equation}
	s \equiv \hat f_\parallel(x,y)\cdot \begin{pmatrix}x\\y\end{pmatrix}~\text{and}~
		s_\perp \equiv \hat f_\perp(x,y)\cdot \begin{pmatrix}x\\y\end{pmatrix}~.
\end{equation}
In order to obtain the dynamics along a noisy limit cycle, we concentrate in the following on the parallel projection. The time-evolution is given by Ito's formula \cite{gardiner} and Eq. (\ref{eq:2DLangevin}) as
\begin{widetext}
\begin{equation}
	\begin{split}
	\frac{d}{dt}s(t) &= \norm{F(x(t),y(t))}  + s_\perp(t) DF(x(t),y(t))\hat f_\parallel(x(t),y(t))\cdot \hat f_\perp(x(t),y(t)) + \mathcal O\left(\frac{1}{\Omega}\right)\\
	 & \quad +\frac{1}{\sqrt \Omega}\hat f_\parallel(x(t),y(t)) \cdot C(x(t),y(t)) \xi(t)  +\frac{1}{\sqrt \Omega}  s_\perp(t) DF(x(t),y(t))\hat f_\parallel(x(t),y(t))\cdot \hat f_\perp(x(t),y(t))\cdot C(x(t),y(t))\xi(t)~.
\end{split}
\label{eq:s}
\end{equation}
\end{widetext}
$DF(x,y)$ denotes the Jacobian matrix of the vector field $F(x,y)$. For briefness of notation, we write $x(t)$ or $y(t)$ in the equation above where we mean $x(t)$ and $y(t)$ parameterized in the new degrees of freedom, i.e., $x(t) = x(s(t),s_\perp(t))$ and similarly for $y(t)$.

In the vicinity of the limit cycle, the dominant direction is parallel to the force. Thus, we neglect the perpendicular component and set $s_\perp(t) = 0$ in Eq. (\ref{eq:s}) to obtain the following Langevin equation
\begin{equation}
	\begin{split}
		\frac{d}{dt} s(t) \simeq F_\parallel(s(t)) + \frac{1}{\sqrt{\Omega}}C_{\parallel}(s(t))\cdot \begin{pmatrix}\xi_1(t) \\\xi_2(t)\end{pmatrix}~
	\end{split}	
	\label{eq:1DLangevin}
\end{equation}
with the force
\begin{equation}
	F_\parallel(s) \equiv \norm{F(x(s,0),y(s,0))}
	\label{eq:Fs}
\end{equation}
and the vector
\begin{equation}
\begin{split}
	C_{\parallel}(s) &\equiv C^T(x(s,0),y(s,0))\cdot \hat f_\parallel(x(s,0),y(s,0)) \\ &\equiv \begin{pmatrix}C_{\parallel,1}(s)\\C_{\parallel,2}(s)\end{pmatrix}~.
\end{split}
	\label{eq:Cs}\end{equation}
Hence, the diffusivity is given by
\begin{equation}
	Q_\parallel(s) \equiv \frac 12 [C_{\parallel,1}(s)^2+C_{\parallel,2}(s)^2]~.
	\label{eq:Qs}\end{equation}
The Ito-Langevin equation, Eq. (\ref{eq:1DLangevin}), is equivalent to the Fokker-Planck equation
\begin{equation}
	\partial_t p(s,t) = -\partial_s[F_\parallel(s)p(s,t)] + \frac{1}{\Omega}\partial^2_s[Q_\parallel(s)p(s,t)]~.
	\label{eq:FP1D}
\end{equation}
We can now use the solution $p(s,t)$ to calculate the auto-correlation function of $x(t)$,
\begin{equation}
	C_x(t) \equiv \braket{x(t)x(0)}~,
\end{equation}
along the noisy trajectory as
\begin{widetext}
\begin{equation}
	\begin{split}
		C_x(t) &= \int dx \int dy \int dx_0 \int dy_0   x x_0 p(x,y,t|x_0,y_0)p_0(x_0,y_0)\\  
		& \simeq \int ds \int ds_\perp  \int ds_{0} \int ds_{\perp,0}   x(s,0)x(s_{0},0) p(s,s_\perp,t|s_{0},s_{\perp,0}) p_0(s_{0}, s_{\perp,0})\\
		& = \int ds \int ds_0 x(s,0) x(s_0,0) p(s,t|s_0)p_0(s_0)\\
		& =\sum_{k\in \mathbb Z}\int ds \int ds_0 x(s)x(s_0)\exp(\lambda^{(k)} t)h^{(k)}(s|s_0)p_0(s_0) \\
		& = 	 C^{(0)} + C^{(1)} \cos(\Im \lambda^{(1)} t) \exp(\Re\lambda^{(1)} t) + \sum_{|k|\geq 2} C^{(k)}\exp(\lambda^{(k)} t)~.
	\end{split}
\end{equation}
\end{widetext}
In the second line, we neglect the $s_\perp(t)$ dependence of $x(t)$ to obtain the behavior parallel to the closed curve. The eigenvalues $\lambda^{(k)}$ and eigenfunctions $h^k(s|s_0)$ are the ones corresponding to the one-dimensional Fokker-Planck equation (\ref{eq:FP1D}) and can be calculated as described above for the one-dimensional system, Sec. \ref{sec:1D}. The constants are given as
\begin{equation}
	C^{(k)} \equiv \int ds \int ds_0 x(s)x(s_0)h^{(k)}(s|s_0)p_0(s_0)~.
\end{equation}

In the weak noise limit, the coordinate $s(t)$ effectively becomes the arc-length. Hence, the integrals appearing in Eq. (\ref{eq:lambda}) can be evaluated according to
\begin{equation}
	\begin{split}
		\int_{s(0)}^{s(T)}\frac{Q_\parallel(s)}{F_\parallel(s)^3} ds& = \int_{0}^{T}\frac{Q_\parallel(\bar x(t),\bar y(t))}{F_\parallel(\bar x(t),\bar y(t))^3}\frac{ds}{dt}  dt \\
		&	=\int_{0}^{T}\frac{Q_\parallel(\bar x(t),\bar y(t))}{F_\parallel(\bar x(t),\bar y(t))^2} dt~.
	\end{split}
\end{equation}
The quality factor of a two-dimensional limit cycle oscillation is thus given by the one-dimensional expression, Eq. (\ref{eq:R1D}), in the weak noise limit with $F_\parallel(s)$ and $Q_\parallel(s)$ as defined above, which is our third main result.

The next two sections are devoted to applications. We study the noisy Stuart-Landau oscillator and discuss the validity of the approximations made above. Afterwards, we examine the thermodynamically consistent Brusselator as a generic model for a chemical clock.

\section{Cubic normal form for Hopf bifurcation}

\subsection{Stuart-Landau oscillator}
\label{sec:examples2}
The cubic normal form for a Hopf bifurcation \cite{stro00, wigg90, frey20,xiao07,bagh14,louc15,tant20a}, often referred to as Stuart-Landau oscillator, is characterized by the following vector field in polar coordinates 
	\begin{equation}
	\begin{split}
		\partial_t r &=  \mu r - a r^3 \equiv F_r(r) \\
		\partial_t \theta &= \omega + b r^2 \equiv F_\theta(r,\theta)~.
	\end{split}
\end{equation}
When $\mu$ changes sign and $a>0$, this field undergoes a Hopf bifurcation and a limit cycle with constant radius $r_0 \equiv \sqrt{\mu/a}$ emerges. We assume the matrix characterizing the noise to be constant,
\begin{equation}
	C(x(t),y(t)) = \sqrt{2 D} \begin{pmatrix}1&0\\0&1\end{pmatrix}~,
\end{equation}
with a free parameter $D>0$.

The normal and tangent degrees of freedom for the closed curve in phase space are the radial deviation $\rho(t)\equiv r(t)-r_0$ and the angle $\theta(t)$, which compare to $s_\perp(t)$ and $s(t)$ from above, respectively. Following the procedure described in Sec. \ref{sec:2D}, we arrive at the Fokker Planck equation for the angle
\begin{equation}
	\partial_t  p(\theta,t) = - \partial_\theta[\omega_0  p(\theta,t)] + \frac{aD}{\mu\Omega} \partial_\theta^2[p(\theta,t)]~.
	\label{eq:FPcubic}
\end{equation}
Thus, $\theta$ follows a free diffusion with constant drift, i.e., 
\begin{equation}
	F_\theta(\theta) \equiv \omega + b \mu/a\equiv \omega_0~\text{~~and~~} Q_\theta(\theta) = aD/\mu~.
	\label{eq:Ftheta}
\end{equation}
 The solution of Eq. (\ref{eq:FPcubic}) is a Gaussian
\begin{equation}
	p(\theta,t|\theta_0) = \frac{1}{\sqrt{4 \pi t Da/(\Omega \mu)}}\exp\bigg[ -\frac 12 \frac{(\theta - \omega_0 t - \theta_0)^2}{2t Da/(\Omega \mu)}\bigg]~.
\end{equation}
For the initial condition $p_0(\theta_0) = \delta(\theta_0 - \theta(0))$, we carry out the Gaussian integrals and obtain
\begin{equation}
	\begin{split}
		C_x(t) &= \frac{\mu}{a} \cos\theta(0)\int d \theta \cos\theta  p(\theta,t|\theta_0)\\ 
		&= \frac{\mu}{a} \cos\theta(0)\int d \theta \frac 12 [\exp(i \theta) + \exp(-i \theta)]p(\theta,t|\theta_0)\\
		& =\frac{\mu}{a} \cos\theta(0)\cos(\omega_0 t + \theta(0))\exp[- (a D / \Omega \mu) t] ~.
	\end{split}
\end{equation}
Thus, we find for the quality factor
\begin{equation}
	\begin{split}
		\mathcal R^{2\textrm d} &=\bigg | \frac{\omega_0}{aD/(\Omega \mu)}\bigg |=\bigg |\frac{\Omega}{D} \frac{\mu}{a} (\omega + b \mu / a)\bigg | = 	\mathcal R_{\mathrm{cont}}
	\end{split}
\label{eq:R2D}
\end{equation}
in agreement with what we get by putting Eq. (\ref{eq:Ftheta}) into the formula for a one-dimensional system, Eq. (\ref{eq:R1D}). Thus, we have analytically shown that the two-dimensional quality factor for the limit cycle oscillation is exactly given by the one-dimensional approximation. Furthermore, since $F_\theta(\theta)$ and $Q_\theta(\theta)$ are constant, $f_{\mathrm{cont}} = \mathcal R_{\mathrm{cont}}$. 

\begin{figure}
	\centering
	\includegraphics[width=.95\textwidth]{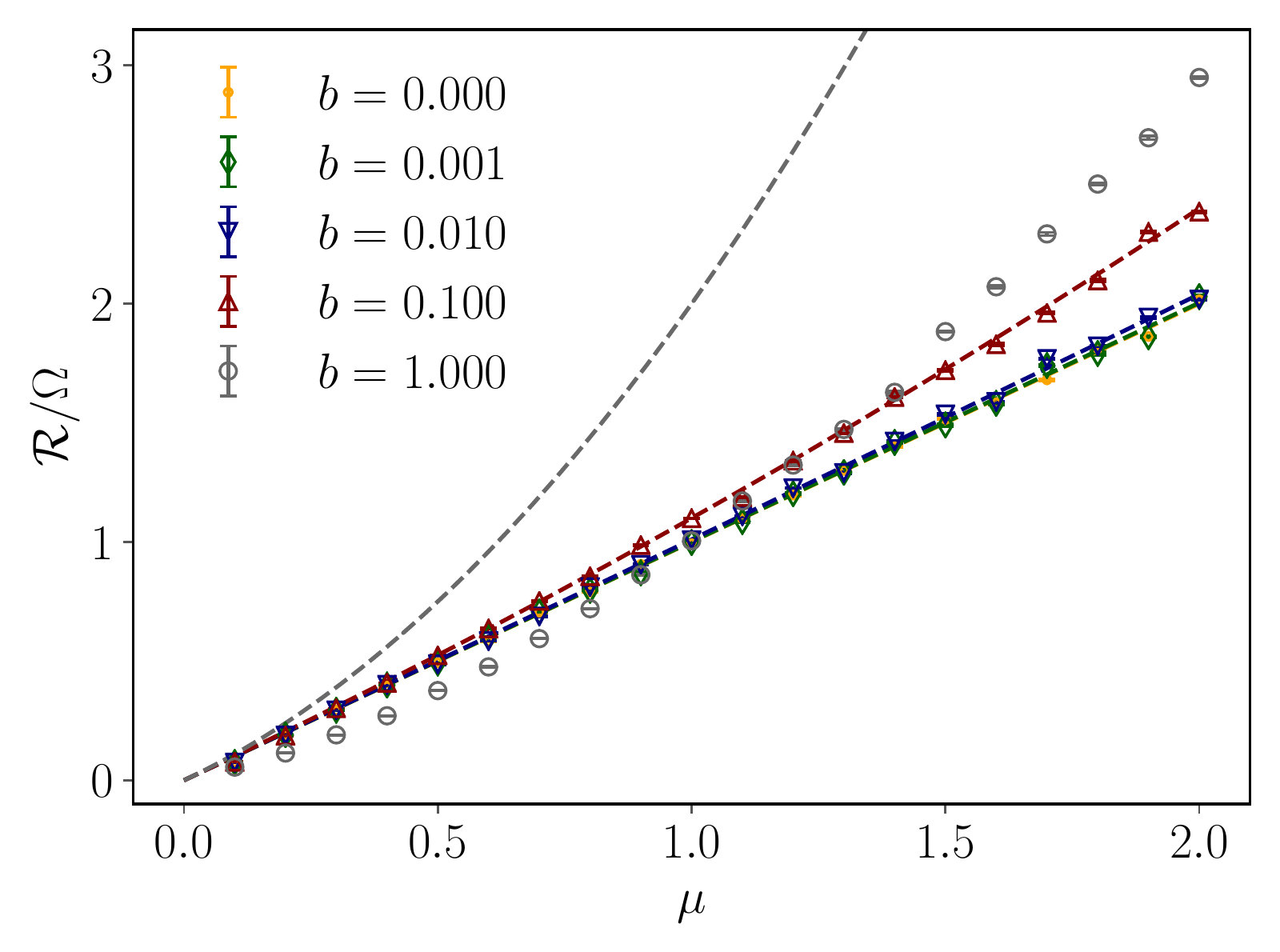}
	\caption{Quality factor for a two-dimensional limit cycle as a function of $\mu$ for different $b$ with $\omega = a = 1$  and $\Omega= 10^3$. Dashed lines represent the theoretical prediction, Eq. (\ref{eq:R2D}). }
	\label{fig:2}
\end{figure}

In Fig. \ref{fig:2}, we compare the quality factor obtained by a simulation of the Langevin dynamics and the analytical result from above. The parameters are chosen such that we obtain a decoupling of the angular and the radial motion, i.e., $|b|\ll 1$. Within this parameter range the numerically obtained quality factor agrees with the analytical one, Eq. (\ref{eq:R2D}). For $b = 1$, we see deviations from the theoretical value. This is due to the contribution of the radial motion to the diffusion of the angle \cite{louc15} as we will show in the next section. 

\subsection{Effective diffusion coefficient}
\label{App:Stuart}

We explain the deviations from the analytical result, Eq. (\ref{eq:R2D}), and show how to expand the prediction into the regime where angular and radial motion are coupled. We start with the solution for the radial deviations $\rho(t) \equiv r(t)-r_0$ of the decoupled Stuart-Landau oscillator.

In leading order for $\Omega \gg 1$, the Ito-Langevin equation is given by 
\begin{equation}
	\begin{split}
		\partial_t \rho(t) & \approx \frac{D}{\Omega}\sqrt{\frac{a}{\mu}} - 2 \mu \rho(t)+ \frac{2D}{\sqrt{\Omega}}[\cos\theta(t) \xi_1(t) + \sin\theta(t)\xi_2(t)]
	\end{split}
	\label{eq:Lrho}\end{equation}
with the corresponding Fokker-Planck equation
\begin{equation}
		\partial_t p(\rho,t) = \partial_{\rho}\bigg[ \bigg(\frac{D}{\Omega}\sqrt{\frac{a}{\mu}} - 2 \mu \rho\bigg)p(\rho,t)\bigg] + \frac{D}{\Omega} \partial^2_{\rho}p(\rho,t)~.
\end{equation}
Thus, the radial deviations follow an Ornstein-Uhlenbeck process with constant drift \cite{louc15}. The mean is given by
\begin{equation}
	\begin{split}
		\braket{\rho(t)} &= \rho_0 \exp(-2\mu t) + \frac{D}{2\Omega} \sqrt{\frac{a}{\mu^3}}[1 - \exp(-2\mu t)] \\ & \approx \rho_0 \exp(-2\mu t),~\text{for}~\Omega \to \infty~,
	\end{split}
\end{equation}
with variance
\begin{equation}
	\begin{split}
		\mathrm{Var}[\rho(t)] & = \frac{D}{2 \Omega \mu} [1 - \exp(-4 \mu t)] \approx  0,~\text{for}~\Omega \to \infty~.
	\end{split}
\end{equation}
The fluctuations of $\rho(t)$ are of order $\Omega^{-1/2}$ and hence, the coupling between $\rho$ and $\theta$ needs to be included if the coupling strength
\begin{equation}
	\partial_r F_\theta(r_0,\theta) = 2b \sqrt{\frac \mu a}
\end{equation}
is not negligible. Since this argument holds for any vector field that does not dependent on $\theta$ in polar coordinates, the following holds beyond the cubic normal form of a Hopf bifurcation. 

\begin{figure}
	\centering
	\includegraphics[width=.95\textwidth]{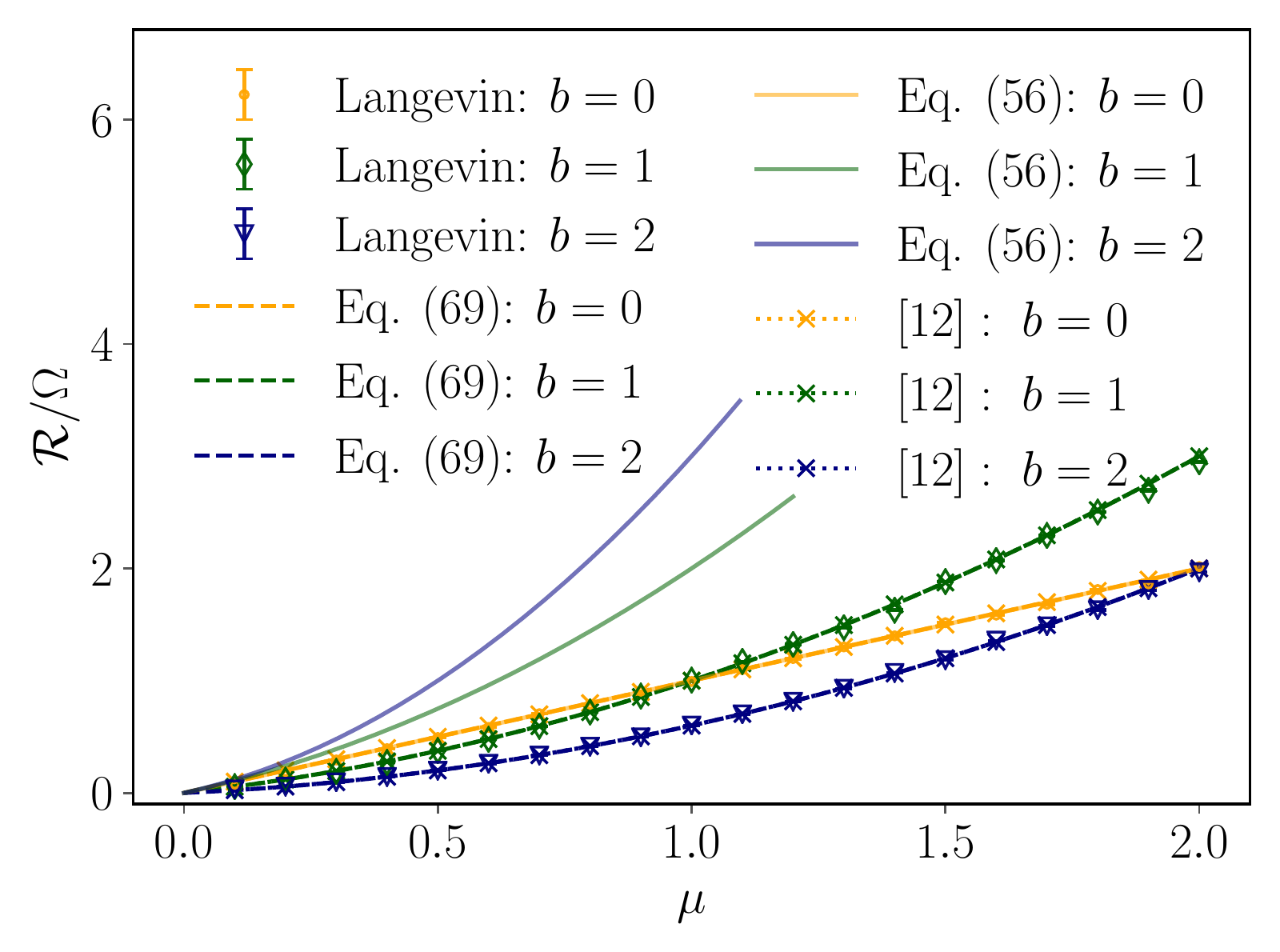}
	\caption{Quality factor $\mathcal R/\Omega$ of the cubic normal form as function of $\mu$ for various $b$. The extensivity-parameter is set to $\Omega = 10^3$ and $\omega = a = 1$. Solid lines represent the one-dimensional prediction, Eq. (\ref{eq:R2D}) and dashed lines the description with the effective diffusion coefficient, Eq. (\ref{eq:Reffective}). Dotted lines with crosses are calculated with the scheme presented in \cite{gasp02}.}
	\label{fig:Appendix1}
\end{figure}

The dynamics of the coupled degrees of freedom is effectively given by the Langevin equations
\begin{equation}
	\begin{split}
		\frac{d}{dt}\begin{pmatrix}\theta(t)\\\rho(t)\end{pmatrix} & = \begin{pmatrix}0 &\kappa_1\\0 & - \kappa_2\end{pmatrix}\begin{pmatrix}\theta(t)\\\rho(t)\end{pmatrix}  +\begin{pmatrix}f_1 \\f_2\end{pmatrix} + \begin{pmatrix} \eta_1(t)\\\eta_2(t)\end{pmatrix}
	\end{split}
\end{equation}
with Gaussian white noise $\braket{\eta_\alpha(t)} = 0$ and $\braket{\eta_\alpha(t)\eta_\beta(t^\prime)} =\sqrt{2D_\alpha} \delta_{\alpha,\beta}\delta(t-t^\prime)$. The $\kappa_\alpha$, $f_\alpha$ and $D_\alpha$ are constants. The "effectively" should be understood such that the equation above leads to the same Fokker-Planck equation as the Langevin equation for $\rho(t)$ and $\theta(t)$ if one chooses the parameters correctly.

The equation for $\rho(t)$ has the solution presented in the beginning of this section. Therefore, we concentrate on $\theta(t)$ and calculate its variance in order to get the diffusion constant. A formal solution is given by
\begin{equation}
	\begin{split}
		\theta(t) &= \theta_0 + \bigg(f_1 + \frac{\kappa_1}{\kappa_2} f_2 \bigg) t + \frac{\kappa_1}{\kappa_2}\bigg( \rho_0 - \frac{f_2}{\kappa_2}\bigg)\bigg(1 - e^{- \kappa_2 t}\bigg)\\
		& \quad +  \int_0^t d\tau \eta_1(\tau) + \frac{\kappa_1}{\kappa_2} \int_0^td\tau \eta_2(\tau)\bigg( 1 - e^{-\kappa_2t}e^{\kappa_2\tau}\bigg)
	\end{split}
\end{equation}
where the integrals are interpreted in the Ito sense and $\theta(0)\equiv \theta_0$ respectively $\rho(0) \equiv \rho_0$. The variance becomes 
\begin{equation}
	\begin{split}
		\braket{\theta(t)^2}-\braket{\theta(t)}^2 &= 2 D_1\bigg(1 + \frac{\kappa_1^2}{\kappa_2^2}\frac{D_2}{D_1}\bigg)~t \\&\quad  + \frac{\kappa_1^2}{\kappa_2^3}D_2 \bigg(1 - e^{-2\kappa_2 t} - 4 + 4 e^{-\kappa_2 t}\bigg)~.
	\end{split}
\end{equation}
We obtain the diffusion coefficient as
\begin{equation}
	\begin{split}
		D_\theta &\equiv \lim_{t\to\infty} \frac{\braket{\theta(t)^2}-\braket{\theta(t)}^2}{2 t}  = \bigg(1 + \frac{\kappa_1^2}{\kappa_2^2}\frac{D_2}{D_1}\bigg) D_1~.
	\end{split}
\end{equation}

For the noisy Stuart-Landau oscillator, we identify the parameters as
\begin{equation}
	\begin{split}
		D_1= \frac{ D a}{\Omega \mu}~, D_2 = \frac{D}{\Omega}~,\kappa_1 = 2 b \sqrt{\frac{\mu}{a}}~, \kappa_2  = 2 \mu~,
\\f_1   = \omega + b \frac{\mu}{a}\equiv \omega_0~, f_2 = \frac{D}{\Omega}\sqrt{\frac{a}{\mu}}~.
	\end{split}
\end{equation}
Thus, the diffusion coefficients becomes
\begin{equation}
D_{\theta} =D_1 \bigg( 1 + \frac{b^2}{a^2}\bigg)
\end{equation}
and the mean angle is given by
\begin{equation}
	\begin{split}
		\braket{\theta(t)} &\approx \theta_0+ \frac{b}{\sqrt{\mu a}} \rho_0+ \bigg(\omega_0 + \frac{b}{\mu}\frac{D}{\Omega}\bigg)~t\\ & = \theta_0+ \frac{b}{\sqrt{\mu a}} \rho_0+ \omega_0~t+ \mathcal O \bigg(\frac 1\Omega \bigg)
	\end{split}
\end{equation}
in the limit of large times. Thus, for large $t$, the angular motion is effectively given by a diffusion with constant $D_{\theta}$ and drift $\omega_0$. According to the calculations in the previous section, the quality factor is then given in the weak noise limit by
\begin{equation}
	\mathcal{\tilde R}^{2\textrm d}= \frac{\omega_0 }{D_\theta}= \frac{ \mathcal R^{2 \textrm d}}{1 + b^2/a^2}~.
	\label{eq:Reffective}
\end{equation}
In Fig. \ref{fig:Appendix1}, we show the quality factor obtained by integrating the Langevin equation, the one-dimensional prediction from the previous section, the effective theory introduced above and data calculated using the method presented in \cite{gasp02}, see Appendix \ref{App:Gaspard} for a brief explanation. In contrast to the previous section, we consider a parameter range with non-negligible coupling between radial deviations and angular motion. As expected, the naive one-dimensional approximation fails to reproduce the quality factor of the underlying Langevin dynamics. Nevertheless, the effective description taking into account radial motion predicts the quality factor correctly and is in agreement with the full theory \cite{gasp02}.

In conclusion, we have extended the validity of the approximations made in Sec. \ref{sec:2D} to a class of systems in which the normal and tangent motion do not necessarily decouple. Moreover, the discussion presented above is not limited to the Stuart-Landau oscillator, it rather holds true for any vector field that does not dependent on the angle. In the next section, we discuss an example for which the latter assumptions does not remain true.

\section{Brusselator}
\label{sec:Brusselator}

A paradigmatic model for a chemical clock is the Brusselator \cite{lefe88,Andrieux2008,nguy18},
\begin{equation}
		A \overset{k_1^+}{\underset{k_1^-}{\rightleftharpoons}} X~,3X	\overset{k_2^-}{\underset{k_2^+}{\rightleftharpoons}} 	2X + Y~,	Y\overset{k_3^-}{\underset{k_3^+}{\rightleftharpoons}}  B~.
\end{equation}
The concentrations $c_A$ and $c_B$ of the chemical species A and B in the external bath are kept constant. Due to a difference in the chemical potential between $A$ and $B$, i.e., $\Delta \mu \equiv \mu_B-\mu_A>0$, the system is out of equilibrium and the number of the intermediate species $X$ and $Y$ can oscillate. Considering the reaction cycle which consumes a substrate $B$ and generates a product $A$, the thermodynamic force associated with this cycle is 
\begin{equation}
	\mathcal A \equiv \Delta \mu = \ln\frac{c_B k_3^+ k_2^+ k_1^-}{c_A k_1 ^+ k_2^-k_3^-}
\label{eq:locDB}
\end{equation}
where $k_i^\pm$ are the corresponding reaction rates. This equation is commonly known as local detailed balance condition \cite{seif12}.

Following \cite{gasp02a}, we obtain $F(x,y)$ and the diffusion matrix $Q(x,y)$ in the weak noise limit as
\begin{equation}
\begin{split}
	F_x(x,y) & = c_A k^+_1 - k^-_1 x + k^+_2 x^2 y - k^-_2 x^3~,\\
	F_y(x,y) & = c_B k^+_3 - k^-_3 y - k^+_2 x^2 y + k^-_2 x^3~,\\
	Q_{x,x}(x,y) & = \frac 12 (c_A k^+_1 + k^-_1 x + k^+_2 x^2 y + k^-_2 x^3)~, \\
	Q_{y,y}(x,y) & = \frac 12 (c_B k^+_3 + k^-_3 y + k^+_2 x^2 y + k^-_2 x^3)~,\\
	Q_{x,y}(x,y) & = Q_{y,x}(x,y) = -\frac 12 (k^+_2 x^2 y + k^-_2 x^3)~,
\end{split}
\end{equation}
where $x \equiv n_X/\Omega$ and $y\equiv n_Y/\Omega$. Here, $n_i$ denotes the number of molecules of species $X$ or $Y$. The external parameter $\Omega$ represents the volume of the system. 

We choose the noise matrix $C(x,y)$, Eq. (\ref{eq:corM}), such that $Q(x,y) = \frac 12 C(x,y)C(x,y)^T$, e.g.,
\begin{equation}
	C(x,y) \equiv \sqrt{\frac{2}{Q_{x,x}(x,y)}}\begin{pmatrix}Q_{x,x}(x,y)& 0\\Q_{x,y}(x,y)& \sqrt{\det Q(x,y)}\end{pmatrix}~.
\end{equation} 
Thus, the noisy rate equation for the Brusselator are of the form of Eq. (\ref{eq:2DLangevin}) and we can apply the theory developed in Sec. \ref{sec:2D}.

The deterministic vector field
\begin{equation}
	F(x,y) \equiv \begin{pmatrix}F_x(x,y)\\F_y(x,y)\end{pmatrix} = \begin{pmatrix}c_A k^+_1 - k^-_1 x + k^+_2 x^2 y - k^-_2 x^3\\
	c_B k^+_3 - k^-_3 y - k^+_2 x^2 y + k^-_2 x^3\\\end{pmatrix}
\end{equation}
 undergoes a Hopf bifurcation while increasing the chemical potential \cite{nguy18}. We obtain the emerging limit cycle by numerically solving the deterministic equation,
 \begin{equation}
 	\frac{d}{dt} \begin{pmatrix} x(t) \\y(t)\end{pmatrix} =  \begin{pmatrix}F_x(x(t),y(t))\\F_y(x(t),y(t))\end{pmatrix}~.
 \end{equation}
Following the procedure from Sec. \ref{sec:2D}, we determine the force $F_\parallel(x,y)$ and diffusivity $Q_\parallel(x,y)$ and numerically obtain our estimate for the quality factor $\mathcal R_{\mathrm{cont}}$.

\begin{figure}
	\subfloat[\label{fig:3}]{%
		\includegraphics[width=.85\textwidth]{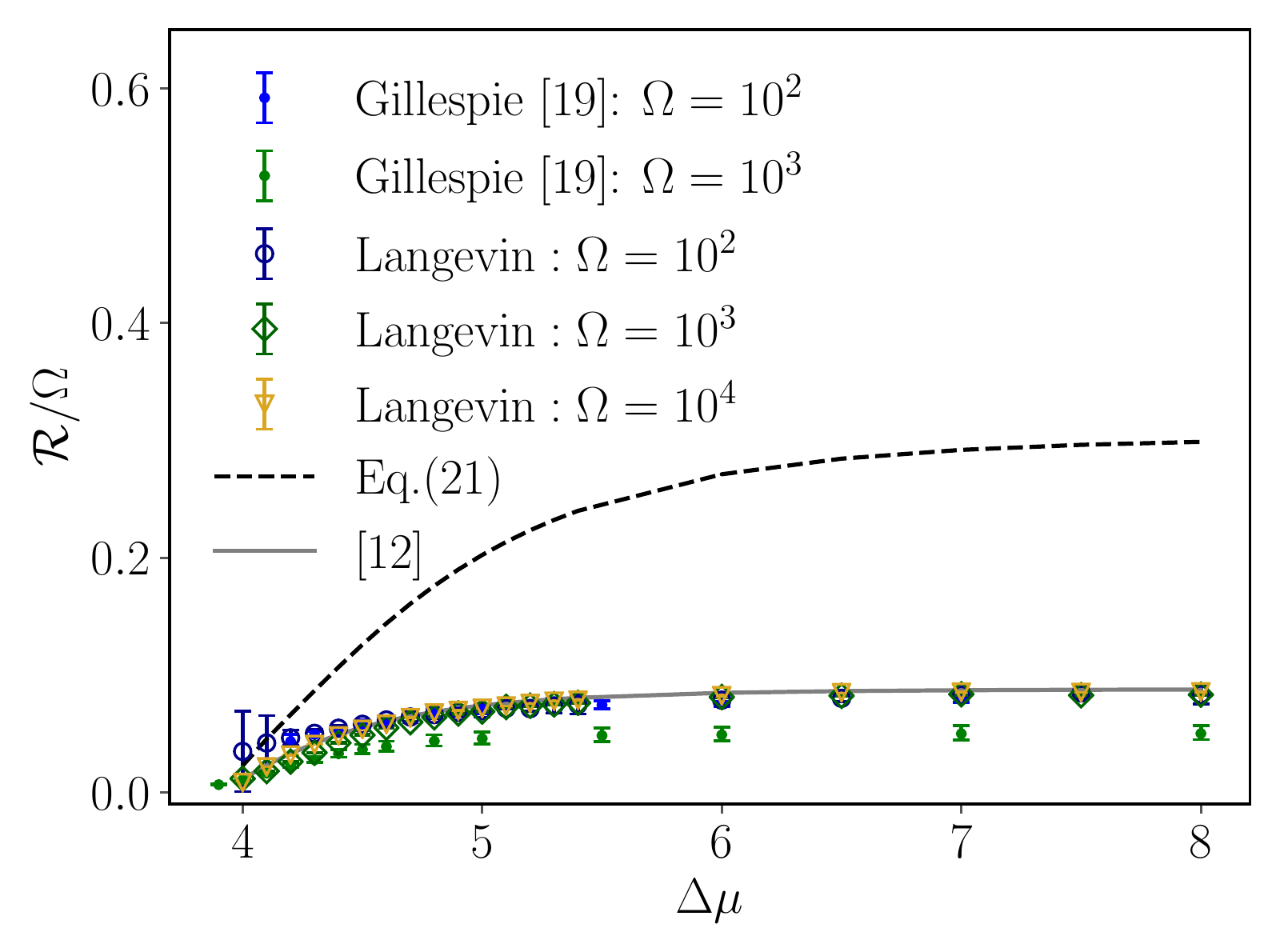}
	}
	
	\subfloat[	\label{fig:4}]{%
		\includegraphics[width=.85\textwidth]{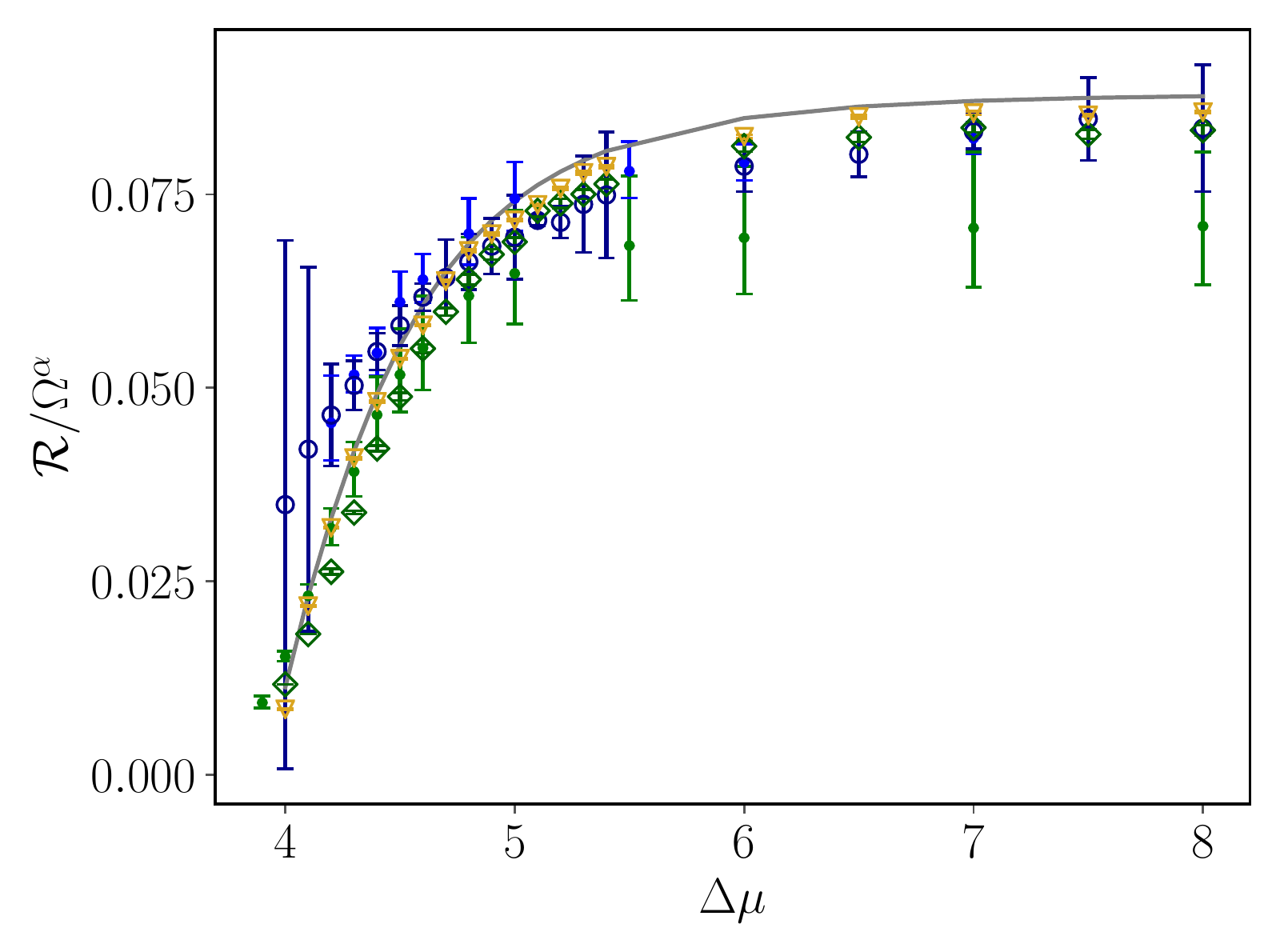}
	}
	\caption{(a) Quality factor and  (b)  $\mathcal R/\Omega^\alpha$ of the Brusselator as function of the thermodynamic force $\Delta \mu$ for various $\Omega$. The rates are chosen as $k_1^+ = k_3^+ = 0.1$, $k_1^- = k_2^+ = k_2^- = 1$, $c_A = 1$, and $c_B = 3$. The rate $k_3^-$ is computed with the thermodynamic force $\Delta\mu $ and relation, Eq. (\ref{eq:locDB}). $\alpha = 0.99$ for the $\Omega = 100$ Gillespie data and  $\alpha = 0.95$ for $\Omega = 10^3$. For all other methods, $\alpha = 1$ by theory. In (b) the prediction by Eq. (\ref{eq:R1D}) is not displayed in order to focus on the collapsed data.}
	\label{fig:Fig3}
\end{figure}

In Fig. \ref{fig:3}, we compare our result with three sets of data; those from \cite{nguy18} as obtained by a Gillespie simulation of the Brusselator, from the algorithm of the full theory \cite{gasp02}, see Appendix \ref{App:Gaspard}, and  from data obtained by a Langevin simulation with the force field $F(x,y)$ and matrix $C(x,y)$ described above. We find for all methods that the quality factor tends to zero in the vicinity of the bifurcation point $\Delta \mu_c\simeq 3.9$ \cite{nguy18} as expected. For $\Delta \mu$ significantly above $\Delta \mu_c$, the data obtained by a Langevin simulation and by the full theory agree, but there are deviations to the the data obtained through the Gillespie algorithm and our method. The variance in the Gillespie data is due to the fact that the number of coherent oscillations, thus the quality factor, scales around the bifurcation-point and away from the transition with an apparent exponent less than 1, i.e., $\mathcal R \propto  \Omega^\alpha$ with $\alpha < 1$ \cite{nguy18}. Re-scaling the data with such an exponent leads to a better agreement with the other data sets, see Fig. \ref{fig:4}. While this procedure leads to consistency with the full theory, there is still a significant discrepancy to the method presented in Sec. \ref{sec:2D}.

\begin{figure}
	\subfloat[	\label{fig:Appendix5}]{%
		\includegraphics[width=.8\textwidth]{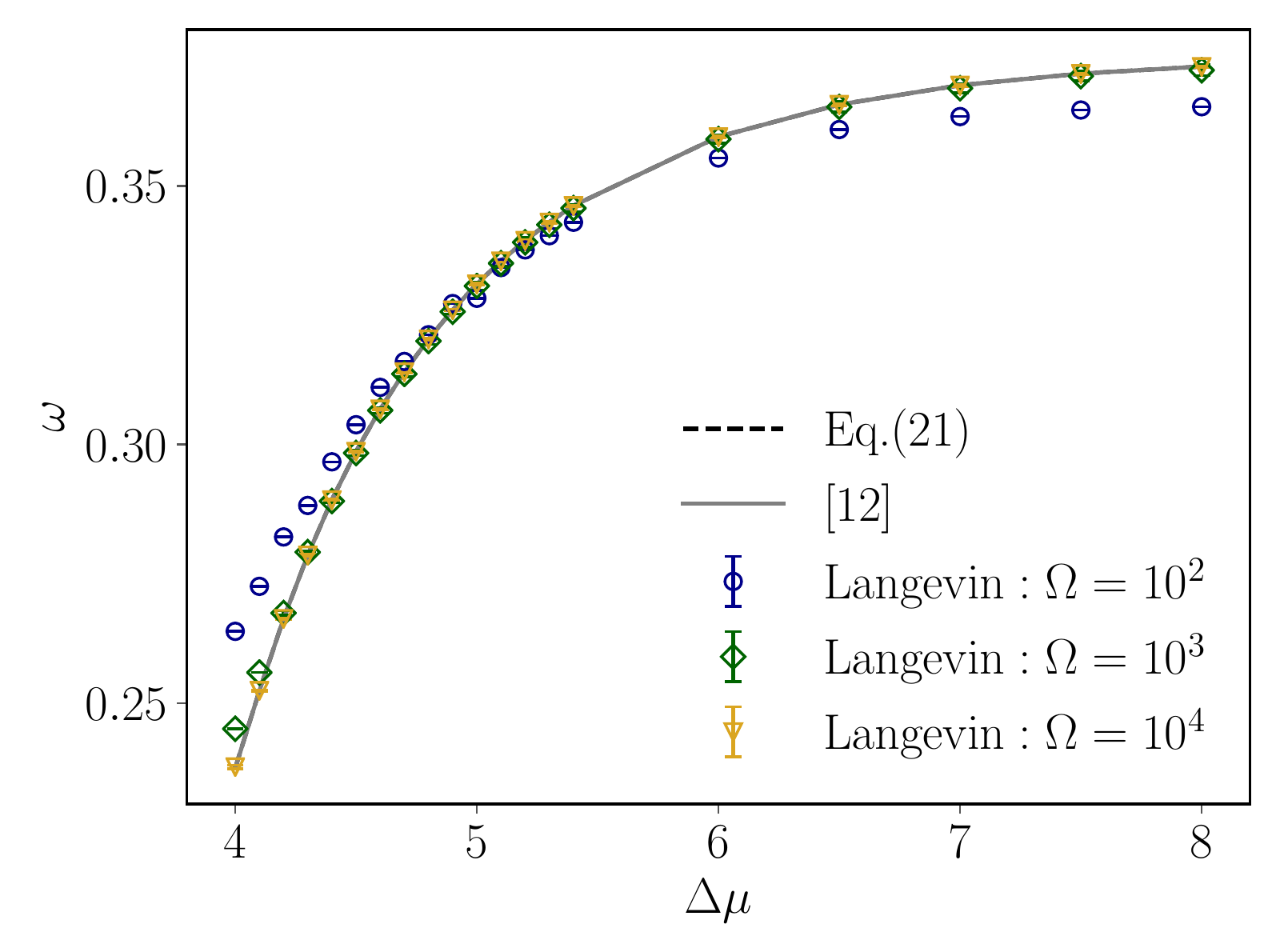}
	}

	\subfloat[	\label{fig:Appendix4}]{%
	\includegraphics[width=.8\textwidth]{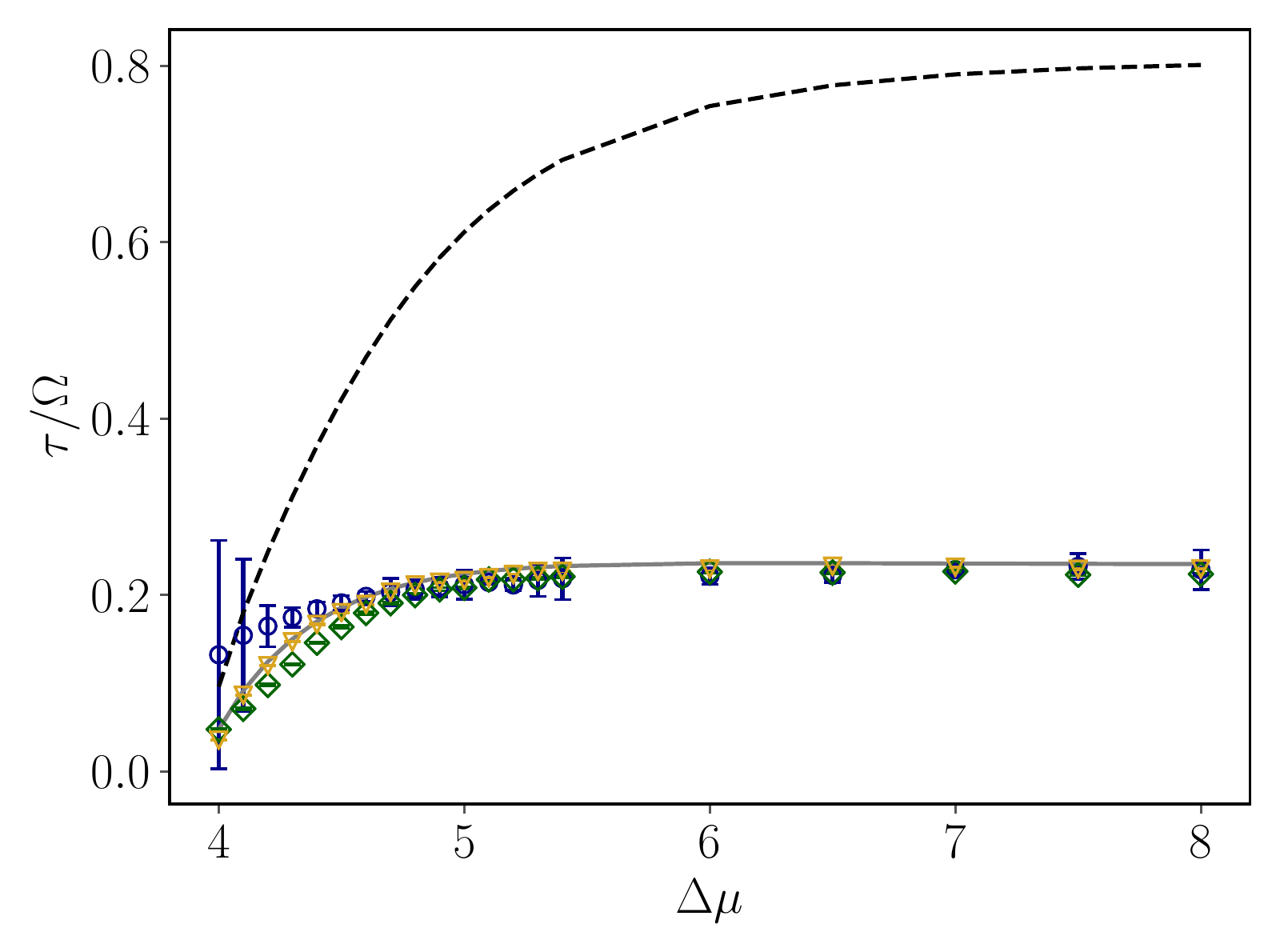}
}
	\caption{(a) Frequency $\omega$  and (b) decay time $\tau/\Omega$  of the Brusselator as function of $\Delta \mu$ for various $\Omega$. The rates are chosen as for Fig. \ref{fig:Fig3}.}
	\label{fig:AppFig4}
\end{figure}

The one-dimensional approximation of the quality factor overestimates the full dynamics of the Brusselator, Fig. \ref{fig:3}. The significant discrepancy between the data obtained through Eq. (\ref{eq:R1D}) and for example by the algorithm \cite{gasp02} resembles the corresponding failure for the noisy Stuart-Landau oscillator, Sec. \ref{App:Stuart}. There, we have analytically shown that the perpendicular motion amplifies the diffusion in the tangent direction. In particular, we have established an effective diffusion constant which sets the decay time of the correlation function. While this analytical treatment is not feasible anymore for the Brusselator due to an angular dependent limit cycle radius, we numerically calculate the frequency $\omega$, Fig. \ref{fig:Appendix5}, and the coherence time $\tau$, Fig. \ref{fig:Appendix4},  for all data sets except the one obtained by the Gillespie algorithm. 

We find only minute variations for the angular frequency. However, the decay time $\tau$ is overestimated by the one-dimensional expression, Eq. (\ref{eq:lambda}), while the remaining methods coincide. Thus, as in the strong coupling regime of the noisy Stuart-Landau oscillator, the diffusion constant for the tangential motion is larger than predicted by the one-dimensional approximation. In fact, this is a generic feature due to the stability of a limit cycle as we have seen in Sec. \ref{App:Stuart}. 

In summary, we have also found for the Brusselator that a coupling between normal and tangent motion leads to a reduced quality factor. Thus, the approximation we have presented in Sec. \ref{sec:2D} can be understood as an effective upper bound on the coherence. For the generic model presented in Sec. \ref{App:Stuart}, this upper bound is sharp.

\section{Discussion and Conclusion}
\label{sec:conclusion}
We have derived an analytical expression for the quality factor for an oscillator based on a one-dimensional Fokker-Planck dynamics, Eq. (\ref{eq:R1D}). This expression is obtained by a spectral decomposition of the Fokker-Planck operator as presented in \cite{gasp02,gasp02a}. In contrast to these studies, we did not use the rather involved techniques of Hamilton-Jacobi theory but solved for the eigenvalues and eigenfunctions directly exploiting the weak-noise limit. Furthermore, we have derived and proven a continuous version of the microscopic bound on the quality factor of a stochastic oscillator which has been conjectured in \cite{bara17} for Markov networks. The continuous bound, Eq. (\ref{eq:BScont}), is tight, since it is derived through the Cauchy-Schwartz inequality, and is simply given by the integral of driving force over diffusion. We have illustrated these results numerically for a particle on a ring subject to a periodic potential and constant diffusion.

This one-dimensional quality factor also captures the behavior of a two-dimensional noisy oscillator. If the coupling of tangent and normal motion is negligible, this correspondence is exact. Moreover, even if the directions do not decouple, the quality factor for a broad class of oscillators can be obtained through an effective one-dimensional expression as we have analytically shown for the Stuart-Landau model as a generic example. The theoretical result is in agreement with data obtained by numerically integrating the corresponding Langevin equation and also matches the numerical data resulting from the method introduced in \cite{gasp02}. This effective description breaks down for systems in which the two-dimensional motion does not decouple. As an example, we  examined the Brusselator numerically and observed that the one-dimensional approximation leads to an overestimate in the decay time as we have also found for the Stuart-Landau oscillator. The full rather involved method introduced in \cite{gasp02,gasp02a} is in good agreement with data obtained by a Gillespie simulation and by integrating the corresponding chemical Langevin equation. Thus, the theory presented in our paper can be understood as an effective upper bound on the quality factor of a two-dimensional noisy oscillator. Indeed, the calculations leading to a parametrization with the arc-length of a limit cycle are not limited to two dimensions but can be generalized to arbitrary dimensions.

In conclusion, we have presented a transparent formula for the quality factor of a noisy oscillator. In one dimension this expression is exact. For a broad class of two-dimensional systems, this expression can be adjusted to yield the correct coherence while for general two-dimensional oscillators, it establishes an effective upper bound. Despite this discrepancy, the method presented in this paper provides an elementary approach to the coherence of a noisy clock. Indeed, for a dynamics in higher dimensions it requires to numerically find the limit cycle and then to integrate a function along the arc-length rather than the numerically more challenging and sensitive techniques required for the full theory \cite{gasp02}.

\begin{acknowledgments}
	We thank B. Nguyen for helpful discussions.
\end{acknowledgments}
\appendix
\section{Irrelevance of the second solution of the one-dimensional Fokker-Planck system}
\label{App:Neglect}
We follow the calculations of Sec. \ref{sec:1D} in the main text and calculate the eigenvalues for the second solution of Eq. (\ref{eq:pert}), 
\begin{equation}
	\tilde g(x) = - \frac{F(x)}{Q(x)} + \frac{1}{\Omega}\bigg(- \frac{\lambda}{F(x)} + \frac{Q^\prime(x)}{Q(x)}\bigg) + \mathcal O \bigg( \frac{1}{\Omega^2}\bigg).
	\label{eqA:g}
\end{equation}
From Eq. (\ref{eq:Periodicity}), we obtain the eigenvalues
\begin{equation}
	\tilde \lambda^{(k)} \approx - i~2 \pi k \frac{1}{\int_0^L 1/F(x) dx} - \Omega \frac{\int_0^L F(x)/Q(x) dx}{\int_0^L 1/F(x) dx}~.
\end{equation}
The resulting quality factor would be given as
\begin{equation}
	 \tilde{\mathcal R} = \bigg | \frac{\Im \tilde \lambda^{(1)}}{\Re \tilde \lambda^{(1)}}\bigg | = \mathcal O \bigg( \frac{1}{\Omega}\bigg),~\text{for}~\Omega \to \infty~.
\end{equation}
Thus, this second solution, Eq. (\ref{eqA:g}), does not contribute to oscillations in the weak noise limit.

%

\section{Method introduced in \cite{gasp02}}
\label{App:Gaspard}
In this Appendix, we briefly discuss the numerical scheme to obtain the quality factor according to \cite{gasp02}. For theoretical background we refer to the original publication \cite{gasp02}, see also \cite{gasp02a,gonz02}.

As first step, one obtains the limit cycle by numerically integrating the deterministic equation of motion, i.e.,
\begin{equation}
\frac{d}{dt} \begin{pmatrix} x(t)\\y(t)\end{pmatrix} = \begin{pmatrix}F_x(x(t),y(t))\\ F_y(x(t),y(t))\end{pmatrix}~.
\end{equation}
We label this periodic solution $(\bar x(t), \bar y(t))$. The period $T$ can be obtained from the numerical data, for instance, by estimating the distance of adjacent maxima.

The next step is to calculate the eigenvectors of the fundamental matrix 
\begin{equation}
	M(T) = \begin{pmatrix} M_{11}(T)& M_{12}(T)\\M_{21}(T) & M_{22}(T)\end{pmatrix}
\end{equation}
at time $T$. In order to do so, one has to solve the following system of differential equations for $0\leq t\leq T$,
\begin{equation}
	\begin{cases}
		\frac{d}{dt} M(t)  = DF(\bar x(t),\bar y(t))~M(t) \\ M(0) = \mathcal{I}
	\end{cases}
\end{equation}
with the two-dimensional unity matrix $\mathcal I$ and the Jacobian matrix $DF(\bar x(t),\bar y(t))$ evaluated along the limit cycle.
The solution $M(T)$ has an eigenvalue $\Lambda_1 = 1$ with right eigenvector 
\begin{equation}
	e_1 \equiv F(\bar x(T),\bar y(T))~,
\end{equation}
which reflects the stability of the periodic solution. The corresponding left eigenvector is denoted by $f_1$, i.e.,
\begin{equation}
M(T)^T~f_1 = f_1~.
\end{equation}
This eigenvector is chosen such that
\begin{equation}
	f_1\cdot e_1 = 1~,
\end{equation}
where we used the standard scalar product as in the main text. The second eigenvalue $\Lambda_2$ with $|\Lambda_2|<1$ and eigenvectors $e_2$ respectively $f_2$ can be ignored.

Deviations from the limit cycle evolve during one period according to the following system of differential equations,
\begin{widetext}
\begin{equation}
	\begin{cases}
\frac{d}{dt}\begin{pmatrix}	\delta  x(t)\\\delta y(t)\end{pmatrix} =\begin{pmatrix} \partial_xF_x(\bar x(t),\bar y(t))& \partial_xF_y(\bar x(t),\bar y(t))\\\partial_yF_x(\bar x(t),\bar y(t)) &\partial_yF_y(\bar x(t),\bar y(t))\end{pmatrix}\begin{pmatrix}	\delta  x(t)\\\delta y(t)\end{pmatrix} &, \begin{pmatrix}	\delta  x(0)\\
	\delta y(0)\end{pmatrix}=\begin{pmatrix}0\\0\end{pmatrix}\\
\frac{d}{dt}\begin{pmatrix}	\delta  p_x(t)\\\delta p_y(t)\end{pmatrix} =- \begin{pmatrix} \partial_xF_x(\bar x(t),\bar y(t))& \partial_xF_y(\bar x(t),\bar y(t))\\\partial_yF_x(\bar x(t),\bar y(t)) &\partial_yF_y(\bar x(t),\bar y(t))\end{pmatrix}^T\begin{pmatrix}	\delta  p_x(t)\\\delta p_y(t)\end{pmatrix}&, \begin{pmatrix}	\delta  p_x(0)\\\delta p_y(0)\end{pmatrix} = \begin{pmatrix}	f_{1,x}\\f_{1,y}\end{pmatrix}.
\end{cases}
\end{equation}
\end{widetext}
As a last step, the solution 
\begin{equation}
	\delta X(T) \equiv \begin{pmatrix}\delta x(T)\\ \delta y(T)\end{pmatrix}
\end{equation} 
determines the quality factor as
\begin{equation}
	\mathcal R \equiv \Omega \frac{T^2}{\pi~|f_1 \cdot \delta X(T)|}~.
\end{equation}

\newpage

\bibliography{../refs2,../../../Bibliography/refs} 

\begin{thebibliography}{39}%
\makeatletter
\providecommand \@ifxundefined [1]{%
 \@ifx{#1\undefined}
}%
\providecommand \@ifnum [1]{%
 \ifnum #1\expandafter \@firstoftwo
 \else \expandafter \@secondoftwo
 \fi
}%
\providecommand \@ifx [1]{%
 \ifx #1\expandafter \@firstoftwo
 \else \expandafter \@secondoftwo
 \fi
}%
\providecommand \natexlab [1]{#1}%
\providecommand \enquote  [1]{``#1''}%
\providecommand \bibnamefont  [1]{#1}%
\providecommand \bibfnamefont [1]{#1}%
\providecommand \citenamefont [1]{#1}%
\providecommand \href@noop [0]{\@secondoftwo}%
\providecommand \href [0]{\begingroup \@sanitize@url \@href}%
\providecommand \@href[1]{\@@startlink{#1}\@@href}%
\providecommand \@@href[1]{\endgroup#1\@@endlink}%
\providecommand \@sanitize@url [0]{\catcode `\\12\catcode `\$12\catcode
  `\&12\catcode `\#12\catcode `\^12\catcode `\_12\catcode `\%12\relax}%
\providecommand \@@startlink[1]{}%
\providecommand \@@endlink[0]{}%
\providecommand \url  [0]{\begingroup\@sanitize@url \@url }%
\providecommand \@url [1]{\endgroup\@href {#1}{\urlprefix }}%
\providecommand \urlprefix  [0]{URL }%
\providecommand \Eprint [0]{\href }%
\providecommand \doibase [0]{http://dx.doi.org/}%
\providecommand \selectlanguage [0]{\@gobble}%
\providecommand \bibinfo  [0]{\@secondoftwo}%
\providecommand \bibfield  [0]{\@secondoftwo}%
\providecommand \translation [1]{[#1]}%
\providecommand \BibitemOpen [0]{}%
\providecommand \bibitemStop [0]{}%
\providecommand \bibitemNoStop [0]{.\EOS\space}%
\providecommand \EOS [0]{\spacefactor3000\relax}%
\providecommand \BibitemShut  [1]{\csname bibitem#1\endcsname}%
\let\auto@bib@innerbib\@empty
\bibitem [{\citenamefont {Ferrell}\ \emph {et~al.}(2011)\citenamefont
  {Ferrell}, \citenamefont {Tsai},\ and\ \citenamefont {Yang}}]{ferr11}%
  \BibitemOpen
  \bibfield  {author} {\bibinfo {author} {\bibfnamefont {James E.}\
  \bibnamefont {Ferrell}}, \bibinfo {author} {\bibfnamefont {Tony Yu-Chen}\
  \bibnamefont {Tsai}}, \ and\ \bibinfo {author} {\bibfnamefont {Qiong}\
  \bibnamefont {Yang}},\ }\bibfield  {title} {\enquote {\bibinfo {title}
  {Modeling the cell cycle: Why do certain circuits oscillate?}}\ }\href
  {\doibase https://doi.org/10.1016/j.cell.2011.03.006} {\bibfield  {journal}
  {\bibinfo  {journal} {Cell}\ }\textbf {\bibinfo {volume} {144}},\ \bibinfo
  {pages} {874--885} (\bibinfo {year} {2011})}\BibitemShut {NoStop}%
\bibitem [{\citenamefont {Masato~Nakajima}(2005)}]{naka05}%
  \BibitemOpen
  \bibfield  {author} {\bibinfo {author} {\bibfnamefont {Hiroshi Ito Taeko
  Nishiwaki Yoriko Murayama Hideo Iwasaki Tokitaka Oyama Takao~Kondo}\
  \bibnamefont {Masato~Nakajima}, \bibfnamefont {Keiko~Imai}},\ }\bibfield
  {title} {\enquote {\bibinfo {title} {Reconstitution of circadian oscillation
  of cyanobacterial kaic phosphorylation in vitro},}\ }\href {\doibase
  10.1126/science.1108451} {\bibfield  {journal} {\bibinfo  {journal}
  {Science}\ }\textbf {\bibinfo {volume} {308}},\ \bibinfo {pages} {414}
  (\bibinfo {year} {2005})}\BibitemShut {NoStop}%
\bibitem [{\citenamefont {Dong}\ and\ \citenamefont {Golden}(2008)}]{dong08}%
  \BibitemOpen
  \bibfield  {author} {\bibinfo {author} {\bibfnamefont {G.}~\bibnamefont
  {Dong}}\ and\ \bibinfo {author} {\bibfnamefont {S.~S.}\ \bibnamefont
  {Golden}},\ }\bibfield  {title} {\enquote {\bibinfo {title} {How a
  cyanobacterium tells time},}\ }\href {\doibase 10.1016/j.mib.2008.10.003}
  {\bibfield  {journal} {\bibinfo  {journal} {Current opinion in microbiology}\
  }\textbf {\bibinfo {volume} {11}},\ \bibinfo {pages} {541--546} (\bibinfo
  {year} {2008})}\BibitemShut {NoStop}%
\bibitem [{\citenamefont {Goldbeter}(1997)}]{gold97}%
  \BibitemOpen
  \bibfield  {author} {\bibinfo {author} {\bibfnamefont {Albert}\ \bibnamefont
  {Goldbeter}},\ }\bibfield  {title} {\enquote {\bibinfo {title} {Modelling
  biochemical oscillations and cellular rhythms},}\ }\href
  {http://www.jstor.org/stable/24101290} {\bibfield  {journal} {\bibinfo
  {journal} {Current Science}\ }\textbf {\bibinfo {volume} {73}},\ \bibinfo
  {pages} {933--939} (\bibinfo {year} {1997})}\BibitemShut {NoStop}%
\bibitem [{\citenamefont {Novák}\ and\ \citenamefont {Tyson}(2008)}]{nova08}%
  \BibitemOpen
  \bibfield  {author} {\bibinfo {author} {\bibfnamefont {Béla}\ \bibnamefont
  {Novák}}\ and\ \bibinfo {author} {\bibfnamefont {John~J}\ \bibnamefont
  {Tyson}},\ }\bibfield  {title} {\enquote {\bibinfo {title} {Design principles
  of biochemical oscillators},}\ }\href {\doibase
  https://doi.org/10.1016/j.cell.2011.03.006} {\bibfield  {journal} {\bibinfo
  {journal} {Nature Reviews Molecular Cell Biology}\ }\textbf {\bibinfo
  {volume} {144}},\ \bibinfo {pages} {874--885} (\bibinfo {year}
  {2008})}\BibitemShut {NoStop}%
\bibitem [{\citenamefont {Nicolis}\ and\ \citenamefont
  {Malek-Mansour}(1978)}]{nico78}%
  \BibitemOpen
  \bibfield  {author} {\bibinfo {author} {\bibfnamefont {G.}~\bibnamefont
  {Nicolis}}\ and\ \bibinfo {author} {\bibfnamefont {M.}~\bibnamefont
  {Malek-Mansour}},\ }\bibfield  {title} {\enquote {\bibinfo {title}
  {{Nonequilibrium Phase Transitions and Chemical Reactions}},}\ }\href
  {\doibase 10.1143/PTPS.64.249} {\bibfield  {journal} {\bibinfo  {journal}
  {Progress of Theoretical Physics Supplement}\ }\textbf {\bibinfo {volume}
  {64}},\ \bibinfo {pages} {249--268} (\bibinfo {year} {1978})},\ \Eprint
  {http://arxiv.org/abs/https://academic.oup.com/ptps/article-pdf/doi/10.1143/PTPS.64.249/5291423/64-249.pdf}
  {https://academic.oup.com/ptps/article-pdf/doi/10.1143/PTPS.64.249/5291423/64-249.pdf}
  \BibitemShut {NoStop}%
\bibitem [{\citenamefont {Strogatz}(2000)}]{stro00}%
  \BibitemOpen
  \bibfield  {author} {\bibinfo {author} {\bibfnamefont {S.}~\bibnamefont
  {Strogatz}},\ }\href@noop {} {\emph {\bibinfo {title} {Nonlinear dynamics and
  chaos: with applications to physics, biology, chemistry and engineering}}}\
  (\bibinfo  {publisher} {Perseus Books Group},\ \bibinfo {year}
  {2000})\BibitemShut {NoStop}%
\bibitem [{\citenamefont {Selkov}(1968)}]{selk68}%
  \BibitemOpen
  \bibfield  {author} {\bibinfo {author} {\bibfnamefont {E.~E.}\ \bibnamefont
  {Selkov}},\ }\bibfield  {title} {\enquote {\bibinfo {title}
  {Self-oscillations in glycolysis 1. a simple kinetic model},}\ }\href
  {\doibase https://doi.org/10.1111/j.1432-1033.1968.tb00175.x} {\bibfield
  {journal} {\bibinfo  {journal} {European Journal of Biochemistry}\ }\textbf
  {\bibinfo {volume} {4}},\ \bibinfo {pages} {79--86} (\bibinfo {year}
  {1968})}\BibitemShut {NoStop}%
\bibitem [{\citenamefont {McQuarrie}(1967)}]{mcqu67}%
  \BibitemOpen
  \bibfield  {author} {\bibinfo {author} {\bibfnamefont {Donald~A.}\
  \bibnamefont {McQuarrie}},\ }\bibfield  {title} {\enquote {\bibinfo {title}
  {Stochastic approach to chemical kinetics},}\ }\href
  {http://www.jstor.org/stable/3212214} {\bibfield  {journal} {\bibinfo
  {journal} {Journal of Applied Probability}\ }\textbf {\bibinfo {volume}
  {4}},\ \bibinfo {pages} {413--478} (\bibinfo {year} {1967})}\BibitemShut
  {NoStop}%
\bibitem [{\citenamefont {{van Kampen}}(1981)}]{vankampen}%
  \BibitemOpen
  \bibfield  {author} {\bibinfo {author} {\bibfnamefont {N.~G.}\ \bibnamefont
  {{van Kampen}}},\ }\href@noop {} {\emph {\bibinfo {title} {Stochastic
  Processes in Physics and Chemistry}}}\ (\bibinfo  {publisher}
  {North-Holland},\ \bibinfo {address} {Amsterdam},\ \bibinfo {year}
  {1981})\BibitemShut {NoStop}%
\bibitem [{\citenamefont {Weber}\ and\ \citenamefont {Frey}(2017)}]{webe17}%
  \BibitemOpen
  \bibfield  {author} {\bibinfo {author} {\bibfnamefont {Markus~F}\
  \bibnamefont {Weber}}\ and\ \bibinfo {author} {\bibfnamefont {Erwin}\
  \bibnamefont {Frey}},\ }\bibfield  {title} {\enquote {\bibinfo {title}
  {Master equations and the theory of stochastic path integrals},}\ }\href
  {\doibase 10.1088/1361-6633/aa5ae2} {\bibfield  {journal} {\bibinfo
  {journal} {Reports on Progress in Physics}\ }\textbf {\bibinfo {volume}
  {80}},\ \bibinfo {pages} {046601} (\bibinfo {year} {2017})}\BibitemShut
  {NoStop}%
\bibitem [{\citenamefont {Gaspard}(2002{\natexlab{a}})}]{gasp02}%
  \BibitemOpen
  \bibfield  {author} {\bibinfo {author} {\bibfnamefont {Pierre}\ \bibnamefont
  {Gaspard}},\ }\bibfield  {title} {\enquote {\bibinfo {title} {{Trace Formula
  for Noisy Flows}},}\ }\href {\doibase {10.1023/A:1013167928166}} {\bibfield
  {journal} {\bibinfo  {journal} {Journal of Statistical Physics}\ }\textbf
  {\bibinfo {volume} {106}},\ \bibinfo {pages} {57--96} (\bibinfo {year}
  {2002}{\natexlab{a}})}\BibitemShut {NoStop}%
\bibitem [{\citenamefont {Gaspard}(2002{\natexlab{b}})}]{gasp02a}%
  \BibitemOpen
  \bibfield  {author} {\bibinfo {author} {\bibfnamefont {Pierre}\ \bibnamefont
  {Gaspard}},\ }\bibfield  {title} {\enquote {\bibinfo {title} {{The
  correlation time of mesoscopic chemical clocks}},}\ }\href {\doibase
  10.1063/1.1513461} {\bibfield  {journal} {\bibinfo  {journal} {Journal of
  Chemical Physics}\ }\textbf {\bibinfo {volume} {117}},\ \bibinfo {pages}
  {8905--8916} (\bibinfo {year} {2002}{\natexlab{b}})}\BibitemShut {NoStop}%
\bibitem [{\citenamefont {Gonze}\ \emph {et~al.}(2002)\citenamefont {Gonze},
  \citenamefont {Halloy},\ and\ \citenamefont {Gaspard}}]{gonz02}%
  \BibitemOpen
  \bibfield  {author} {\bibinfo {author} {\bibfnamefont {D.}~\bibnamefont
  {Gonze}}, \bibinfo {author} {\bibfnamefont {J.}~\bibnamefont {Halloy}}, \
  and\ \bibinfo {author} {\bibfnamefont {P.}~\bibnamefont {Gaspard}},\
  }\bibfield  {title} {\enquote {\bibinfo {title} {{Biochemical clocks and
  molecular noise: Theoretical study of robustness factors}},}\ }\href
  {\doibase 10.1063/1.1475765} {\bibfield  {journal} {\bibinfo  {journal}
  {Journal of Chemical Physics}\ }\textbf {\bibinfo {volume} {116}},\ \bibinfo
  {pages} {10997--11010} (\bibinfo {year} {2002})}\BibitemShut {NoStop}%
\bibitem [{\citenamefont {Qian}(2006)}]{qian06}%
  \BibitemOpen
  \bibfield  {author} {\bibinfo {author} {\bibfnamefont {H.}~\bibnamefont
  {Qian}},\ }\bibfield  {title} {\enquote {\bibinfo {title} {Open-system
  nonequilibrium steady state: Statistical thermodynamics, fluctuations, and
  chemical oscillations},}\ }\href {\doibase 10.1021/jp061858z} {\bibfield
  {journal} {\bibinfo  {journal} {J.\ Phys.\ Chem.\ B}\ }\textbf {\bibinfo
  {volume} {110}},\ \bibinfo {pages} {15063--15074} (\bibinfo {year}
  {2006})}\BibitemShut {NoStop}%
\bibitem [{\citenamefont {Morelli}\ and\ \citenamefont
  {J\"ulicher}(2007)}]{more07}%
  \BibitemOpen
  \bibfield  {author} {\bibinfo {author} {\bibfnamefont {Luis~G.}\ \bibnamefont
  {Morelli}}\ and\ \bibinfo {author} {\bibfnamefont {Frank}\ \bibnamefont
  {J\"ulicher}},\ }\bibfield  {title} {\enquote {\bibinfo {title} {Precision of
  genetic oscillators and clocks},}\ }\href {\doibase
  10.1103/PhysRevLett.98.228101} {\bibfield  {journal} {\bibinfo  {journal}
  {Phys. Rev. Lett.}\ }\textbf {\bibinfo {volume} {98}},\ \bibinfo {pages}
  {228101} (\bibinfo {year} {2007})}\BibitemShut {NoStop}%
\bibitem [{\citenamefont {d{\textquotesingle}Eysmond}\ \emph
  {et~al.}(2013)\citenamefont {d{\textquotesingle}Eysmond}, \citenamefont
  {Simone},\ and\ \citenamefont {Naef}}]{deys13}%
  \BibitemOpen
  \bibfield  {author} {\bibinfo {author} {\bibfnamefont {Thomas}\ \bibnamefont
  {d{\textquotesingle}Eysmond}}, \bibinfo {author} {\bibfnamefont
  {Alessandro~De}\ \bibnamefont {Simone}}, \ and\ \bibinfo {author}
  {\bibfnamefont {Felix}\ \bibnamefont {Naef}},\ }\bibfield  {title} {\enquote
  {\bibinfo {title} {Analysis of precision in chemical oscillators:
  implications for circadian clocks},}\ }\href {\doibase
  10.1088/1478-3975/10/5/056005} {\bibfield  {journal} {\bibinfo  {journal}
  {Physical Biology}\ }\textbf {\bibinfo {volume} {10}},\ \bibinfo {pages}
  {056005} (\bibinfo {year} {2013})}\BibitemShut {NoStop}%
\bibitem [{\citenamefont {Barato}\ and\ \citenamefont
  {Seifert}(2017)}]{bara17}%
  \BibitemOpen
  \bibfield  {author} {\bibinfo {author} {\bibfnamefont {Andre~C.}\
  \bibnamefont {Barato}}\ and\ \bibinfo {author} {\bibfnamefont {Udo}\
  \bibnamefont {Seifert}},\ }\bibfield  {title} {\enquote {\bibinfo {title}
  {Coherence of biochemical oscillations is bounded by driving force and
  network topology},}\ }\href {\doibase 10.1103/PhysRevE.95.062409} {\bibfield
  {journal} {\bibinfo  {journal} {Phys. Rev. E}\ }\textbf {\bibinfo {volume}
  {95}},\ \bibinfo {pages} {062409} (\bibinfo {year} {2017})}\BibitemShut
  {NoStop}%
\bibitem [{\citenamefont {Nguyen}\ \emph {et~al.}(2018)\citenamefont {Nguyen},
  \citenamefont {Seifert},\ and\ \citenamefont {Barato}}]{nguy18}%
  \BibitemOpen
  \bibfield  {author} {\bibinfo {author} {\bibfnamefont {Basile}\ \bibnamefont
  {Nguyen}}, \bibinfo {author} {\bibfnamefont {Udo}\ \bibnamefont {Seifert}}, \
  and\ \bibinfo {author} {\bibfnamefont {Andre~C.}\ \bibnamefont {Barato}},\
  }\bibfield  {title} {\enquote {\bibinfo {title} {{Phase transition in
  thermodynamically consistent biochemical oscillators}},}\ }\href {\doibase
  10.1063/1.5032104} {\bibfield  {journal} {\bibinfo  {journal} {Journal of
  Chemical Physics}\ }\textbf {\bibinfo {volume} {149}} (\bibinfo {year}
  {2018}),\ 10.1063/1.5032104},\ \Eprint {http://arxiv.org/abs/1804.01080}
  {arXiv:1804.01080} \BibitemShut {NoStop}%
\bibitem [{\citenamefont {Barato}\ and\ \citenamefont
  {Seifert}(2016)}]{bara16}%
  \BibitemOpen
  \bibfield  {author} {\bibinfo {author} {\bibfnamefont {Andre~C.}\
  \bibnamefont {Barato}}\ and\ \bibinfo {author} {\bibfnamefont {Udo}\
  \bibnamefont {Seifert}},\ }\bibfield  {title} {\enquote {\bibinfo {title}
  {Cost and precision of {B}rownian clocks},}\ }\href {\doibase
  10.1103/PhysRevX.6.041053} {\bibfield  {journal} {\bibinfo  {journal} {Phys.
  Rev. X}\ }\textbf {\bibinfo {volume} {6}},\ \bibinfo {pages} {041053}
  (\bibinfo {year} {2016})}\BibitemShut {NoStop}%
\bibitem [{\citenamefont {Laurent Potvin-Trottier}\ and\ \citenamefont
  {Paulsson}(2016)}]{potv16}%
  \BibitemOpen
  \bibfield  {author} {\bibinfo {author} {\bibfnamefont {Glenn~Vinnicombe}\
  \bibnamefont {Laurent Potvin-Trottier}, \bibfnamefont {Nathan D.~Lord}}\ and\
  \bibinfo {author} {\bibfnamefont {Johan}\ \bibnamefont {Paulsson}},\
  }\bibfield  {title} {\enquote {\bibinfo {title} {Synchronous long-term
  oscillations in a synthetic gene circuit},}\ }\href {\doibase
  10.1038/nature19841} {\bibfield  {journal} {\bibinfo  {journal} {Nature}\
  }\textbf {\bibinfo {volume} {538}},\ \bibinfo {pages} {514--517} (\bibinfo
  {year} {2016})}\BibitemShut {NoStop}%
\bibitem [{\citenamefont {Barato}\ and\ \citenamefont
  {Seifert}(2015)}]{bara15}%
  \BibitemOpen
  \bibfield  {author} {\bibinfo {author} {\bibfnamefont {Andre~C.}\
  \bibnamefont {Barato}}\ and\ \bibinfo {author} {\bibfnamefont {Udo}\
  \bibnamefont {Seifert}},\ }\bibfield  {title} {\enquote {\bibinfo {title}
  {Thermodynamic uncertainty relation for biomolecular processes},}\ }\href
  {\doibase 10.1103/PhysRevLett.114.158101} {\bibfield  {journal} {\bibinfo
  {journal} {Phys.\ Rev.\ Lett.}\ }\textbf {\bibinfo {volume} {114}},\ \bibinfo
  {pages} {158101} (\bibinfo {year} {2015})}\BibitemShut {NoStop}%
\bibitem [{\citenamefont {Gingrich}\ \emph {et~al.}(2016)\citenamefont
  {Gingrich}, \citenamefont {Horowitz}, \citenamefont {Perunov},\ and\
  \citenamefont {England}}]{ging16}%
  \BibitemOpen
  \bibfield  {author} {\bibinfo {author} {\bibfnamefont {Todd~R.}\ \bibnamefont
  {Gingrich}}, \bibinfo {author} {\bibfnamefont {Jordan~M.}\ \bibnamefont
  {Horowitz}}, \bibinfo {author} {\bibfnamefont {Nikolay}\ \bibnamefont
  {Perunov}}, \ and\ \bibinfo {author} {\bibfnamefont {Jeremy~L.}\ \bibnamefont
  {England}},\ }\bibfield  {title} {\enquote {\bibinfo {title} {Dissipation
  bounds all steady-state current fluctuations},}\ }\href {\doibase
  10.1103/PhysRevLett.116.120601} {\bibfield  {journal} {\bibinfo  {journal}
  {Phys.\ Rev.\ Lett.}\ }\textbf {\bibinfo {volume} {116}},\ \bibinfo {pages}
  {120601} (\bibinfo {year} {2016})}\BibitemShut {NoStop}%
\bibitem [{\citenamefont {Horowitz}\ and\ \citenamefont
  {Gingrich}(2020)}]{horo20}%
  \BibitemOpen
  \bibfield  {author} {\bibinfo {author} {\bibfnamefont {Jordan~M.}\
  \bibnamefont {Horowitz}}\ and\ \bibinfo {author} {\bibfnamefont {Todd~R.}\
  \bibnamefont {Gingrich}},\ }\bibfield  {title} {\enquote {\bibinfo {title}
  {Thermodynamic uncertainty relations constrain non-equilibrium
  fluctuations},}\ }\href {\doibase https://doi.org/10.1038/s41567-019-0702-6}
  {\bibfield  {journal} {\bibinfo  {journal} {Nat. Phys.}\ }\textbf {\bibinfo
  {volume} {16}},\ \bibinfo {pages} {15--20} (\bibinfo {year}
  {2020})}\BibitemShut {NoStop}%
\bibitem [{\citenamefont {Pietzonka}(2021)}]{piet21}%
  \BibitemOpen
  \bibfield  {author} {\bibinfo {author} {\bibfnamefont {Patrick}\ \bibnamefont
  {Pietzonka}},\ }\href@noop {} {\enquote {\bibinfo {title} {Classical pendulum
  clocks break the thermodynamic uncertainty relation},}\ } (\bibinfo {year}
  {2021}),\ \Eprint {http://arxiv.org/abs/2110.02213} {arXiv:2110.02213
  [cond-mat.stat-mech]} \BibitemShut {NoStop}%
\bibitem [{\citenamefont {Freitas}\ \emph {et~al.}(2021)\citenamefont
  {Freitas}, \citenamefont {Falasco},\ and\ \citenamefont {Esposito}}]{frei21}%
  \BibitemOpen
  \bibfield  {author} {\bibinfo {author} {\bibfnamefont {Nahuel}\ \bibnamefont
  {Freitas}}, \bibinfo {author} {\bibfnamefont {Gianmaria}\ \bibnamefont
  {Falasco}}, \ and\ \bibinfo {author} {\bibfnamefont {Massimiliano}\
  \bibnamefont {Esposito}},\ }\bibfield  {title} {\enquote {\bibinfo {title}
  {Linear response in large deviations theory: a method to compute
  non-equilibrium distributions},}\ }\href {\doibase 10.1088/1367-2630/ac1bf5}
  {\bibfield  {journal} {\bibinfo  {journal} {New Journal of Physics}\ }\textbf
  {\bibinfo {volume} {23}},\ \bibinfo {pages} {093003} (\bibinfo {year}
  {2021})}\BibitemShut {NoStop}%
\bibitem [{\citenamefont {Yoshimura}\ and\ \citenamefont
  {Arai}(2008)}]{yosh08}%
  \BibitemOpen
  \bibfield  {author} {\bibinfo {author} {\bibfnamefont {Kazuyuki}\
  \bibnamefont {Yoshimura}}\ and\ \bibinfo {author} {\bibfnamefont {Kenichi}\
  \bibnamefont {Arai}},\ }\bibfield  {title} {\enquote {\bibinfo {title} {Phase
  reduction of stochastic limit cycle oscillators},}\ }\href {\doibase
  10.1103/PhysRevLett.101.154101} {\bibfield  {journal} {\bibinfo  {journal}
  {Phys. Rev. Lett.}\ }\textbf {\bibinfo {volume} {101}},\ \bibinfo {pages}
  {154101} (\bibinfo {year} {2008})}\BibitemShut {NoStop}%
\bibitem [{\citenamefont {Teramae}\ \emph {et~al.}(2009)\citenamefont
  {Teramae}, \citenamefont {Nakao},\ and\ \citenamefont {Ermentrout}}]{junn09}%
  \BibitemOpen
  \bibfield  {author} {\bibinfo {author} {\bibfnamefont {Jun-nosuke}\
  \bibnamefont {Teramae}}, \bibinfo {author} {\bibfnamefont {Hiroya}\
  \bibnamefont {Nakao}}, \ and\ \bibinfo {author} {\bibfnamefont {G.B.}\
  \bibnamefont {Ermentrout}},\ }\bibfield  {title} {\enquote {\bibinfo {title}
  {Stochastic phase reduction for a general class of noisy limit cycle
  oscillators},}\ }\href {\doibase 10.1103/PhysRevLett.102.194102} {\bibfield
  {journal} {\bibinfo  {journal} {Phys. Rev. Lett.}\ }\textbf {\bibinfo
  {volume} {102}},\ \bibinfo {pages} {194102} (\bibinfo {year}
  {2009})}\BibitemShut {NoStop}%
\bibitem [{\citenamefont {Cao}\ \emph {et~al.}(2015)\citenamefont {Cao},
  \citenamefont {Wang}, \citenamefont {Ouyang},\ and\ \citenamefont
  {Tu}}]{cao15}%
  \BibitemOpen
  \bibfield  {author} {\bibinfo {author} {\bibfnamefont {Yuansheng}\
  \bibnamefont {Cao}}, \bibinfo {author} {\bibfnamefont {Hongli}\ \bibnamefont
  {Wang}}, \bibinfo {author} {\bibfnamefont {Qi}~\bibnamefont {Ouyang}}, \ and\
  \bibinfo {author} {\bibfnamefont {Yuhai}\ \bibnamefont {Tu}},\ }\bibfield
  {title} {\enquote {\bibinfo {title} {The free-energy cost of accurate
  biochemical oscillations},}\ }\href {\doibase 10.1038/nphys3412} {\bibfield
  {journal} {\bibinfo  {journal} {Nature Phys.}\ }\textbf {\bibinfo {volume}
  {11}},\ \bibinfo {pages} {772--778} (\bibinfo {year} {2015})}\BibitemShut
  {NoStop}%
\bibitem [{\citenamefont {Bagheri}(2014)}]{bagh14}%
  \BibitemOpen
  \bibfield  {author} {\bibinfo {author} {\bibfnamefont {Shervin}\ \bibnamefont
  {Bagheri}},\ }\bibfield  {title} {\enquote {\bibinfo {title} {{Effects of
  weak noise on oscillating flows: Linking quality factor, floquet modes, and
  koopman spectrum}},}\ }\href {\doibase 10.1063/1.4895898} {\bibfield
  {journal} {\bibinfo  {journal} {Physics of Fluids}\ }\textbf {\bibinfo
  {volume} {26}} (\bibinfo {year} {2014}),\ 10.1063/1.4895898}\BibitemShut
  {NoStop}%
\bibitem [{\citenamefont {Gardiner}(2004)}]{gardiner}%
  \BibitemOpen
  \bibfield  {author} {\bibinfo {author} {\bibfnamefont {C.~W.}\ \bibnamefont
  {Gardiner}},\ }\href@noop {} {\emph {\bibinfo {title} {Handbook of Stochastic
  Methods}}},\ \bibinfo {edition} {3rd}\ ed.\ (\bibinfo  {publisher}
  {Springer-Verlag},\ \bibinfo {address} {Berlin},\ \bibinfo {year}
  {2004})\BibitemShut {NoStop}%
\bibitem [{\citenamefont {Wiggins}(1990)}]{wigg90}%
  \BibitemOpen
  \bibinfo {editor} {\bibfnamefont {Stephen}\ \bibnamefont {Wiggins}},\ ed.,\
  \href@noop {} {\emph {\bibinfo {title} {Introduction to applied nonlinear
  dynamical systems and chaos}}},\ \bibinfo {edition} {2nd}\ ed.,\ Texts in
  applied mathematics ; 2\ (\bibinfo  {publisher} {Springer},\ \bibinfo
  {address} {New York ; Berlin ; Heidelberg},\ \bibinfo {year}
  {1990})\BibitemShut {NoStop}%
\bibitem [{\citenamefont {Frey}\ and\ \citenamefont {Brauns}(2020)}]{frey20}%
  \BibitemOpen
  \bibfield  {author} {\bibinfo {author} {\bibfnamefont {Erwin}\ \bibnamefont
  {Frey}}\ and\ \bibinfo {author} {\bibfnamefont {Fridtjof}\ \bibnamefont
  {Brauns}},\ }\href@noop {} {\enquote {\bibinfo {title} {Self-organisation of
  protein patterns},}\ } (\bibinfo {year} {2020}),\ \Eprint
  {http://arxiv.org/abs/2012.01797} {arXiv:2012.01797 [physics.bio-ph]}
  \BibitemShut {NoStop}%
\bibitem [{\citenamefont {Xiao}\ \emph {et~al.}(2007)\citenamefont {Xiao},
  \citenamefont {Ma}, \citenamefont {Hou},\ and\ \citenamefont {Xin}}]{xiao07}%
  \BibitemOpen
  \bibfield  {author} {\bibinfo {author} {\bibfnamefont {T.~J.}\ \bibnamefont
  {Xiao}}, \bibinfo {author} {\bibfnamefont {J.}~\bibnamefont {Ma}}, \bibinfo
  {author} {\bibfnamefont {Z.~H.}\ \bibnamefont {Hou}}, \ and\ \bibinfo
  {author} {\bibfnamefont {H.}~\bibnamefont {Xin}},\ }\bibfield  {title}
  {\enquote {\bibinfo {title} {Effects of internal noise in mesoscopic chemical
  systems near hopf bifurcation},}\ }\href {\doibase
  10.1088/1367-2630/9/11/403} {\bibfield  {journal} {\bibinfo  {journal} {New
  J. Phys.}\ }\textbf {\bibinfo {volume} {9}},\ \bibinfo {pages} {403}
  (\bibinfo {year} {2007})}\BibitemShut {NoStop}%
\bibitem [{\citenamefont {Louca}(2018)}]{louc15}%
  \BibitemOpen
  \bibfield  {author} {\bibinfo {author} {\bibfnamefont {Stilianos}\
  \bibnamefont {Louca}},\ }\href@noop {} {\enquote {\bibinfo {title} {Stable
  limit cycles perturbed by noise},}\ } (\bibinfo {year} {2018}),\ \Eprint
  {http://arxiv.org/abs/1506.00756} {arXiv:1506.00756 [math.DS]} \BibitemShut
  {NoStop}%
\bibitem [{\citenamefont {Tantet}\ \emph {et~al.}(2020)\citenamefont {Tantet},
  \citenamefont {Chekroun}, \citenamefont {Neelin},\ and\ \citenamefont
  {Dijkstra}}]{tant20a}%
  \BibitemOpen
  \bibfield  {author} {\bibinfo {author} {\bibfnamefont {Alexis}\ \bibnamefont
  {Tantet}}, \bibinfo {author} {\bibfnamefont {Micka{\"{e}}l~D.}\ \bibnamefont
  {Chekroun}}, \bibinfo {author} {\bibfnamefont {J.~David}\ \bibnamefont
  {Neelin}}, \ and\ \bibinfo {author} {\bibfnamefont {Henk~A.}\ \bibnamefont
  {Dijkstra}},\ }\bibfield  {title} {\enquote {\bibinfo {title}
  {{Ruelle–Pollicott Resonances of Stochastic Systems in Reduced State Space.
  Part III: Application to the Cane–Zebiak Model of the El
  Ni{\~{n}}o–Southern Oscillation}},}\ }\href {\doibase
  10.1007/s10955-019-02444-8} {\bibfield  {journal} {\bibinfo  {journal}
  {Journal of Statistical Physics}\ }\textbf {\bibinfo {volume} {179}},\
  \bibinfo {pages} {1449--1474} (\bibinfo {year} {2020})}\BibitemShut {NoStop}%
\bibitem [{\citenamefont {Lefever}\ \emph {et~al.}(1988)\citenamefont
  {Lefever}, \citenamefont {Nicolis},\ and\ \citenamefont
  {Borckmans}}]{lefe88}%
  \BibitemOpen
  \bibfield  {author} {\bibinfo {author} {\bibfnamefont {René}\ \bibnamefont
  {Lefever}}, \bibinfo {author} {\bibfnamefont {Grégoire}\ \bibnamefont
  {Nicolis}}, \ and\ \bibinfo {author} {\bibfnamefont {Pierre}\ \bibnamefont
  {Borckmans}},\ }\bibfield  {title} {\enquote {\bibinfo {title} {The
  brusselator: it does oscillate all the same},}\ }\href {\doibase
  10.1039/F19888401013} {\bibfield  {journal} {\bibinfo  {journal} {J. Chem.
  Soc.{,} Faraday Trans. 1}\ }\textbf {\bibinfo {volume} {84}},\ \bibinfo
  {pages} {1013--1023} (\bibinfo {year} {1988})}\BibitemShut {NoStop}%
\bibitem [{\citenamefont {Andrieux}\ and\ \citenamefont
  {Gaspard}(2008)}]{Andrieux2008}%
  \BibitemOpen
  \bibfield  {author} {\bibinfo {author} {\bibfnamefont {David}\ \bibnamefont
  {Andrieux}}\ and\ \bibinfo {author} {\bibfnamefont {Pierre}\ \bibnamefont
  {Gaspard}},\ }\bibfield  {title} {\enquote {\bibinfo {title} {{Fluctuation
  theorem and mesoscopic chemical clocks}},}\ }\href {\doibase
  10.1063/1.2894475} {\bibfield  {journal} {\bibinfo  {journal} {Journal of
  Chemical Physics}\ }\textbf {\bibinfo {volume} {128}} (\bibinfo {year}
  {2008}),\ 10.1063/1.2894475}\BibitemShut {NoStop}%
\bibitem [{\citenamefont {Seifert}(2012)}]{seif12}%
  \BibitemOpen
  \bibfield  {author} {\bibinfo {author} {\bibfnamefont {U.}~\bibnamefont
  {Seifert}},\ }\bibfield  {title} {\enquote {\bibinfo {title} {Stochastic
  thermodynamics, fluctuation theorems, and molecular machines},}\ }\href
  {\doibase 10.1088/0034-4885/75/12/126001} {\bibfield  {journal} {\bibinfo
  {journal} {Rep. Prog. Phys.}\ }\textbf {\bibinfo {volume} {75}},\ \bibinfo
  {pages} {126001} (\bibinfo {year} {2012})}\BibitemShut {NoStop}%
\end{thebibliography}%

\end{document}